\documentclass[%
 reprint,
 amsmath,amssymb,
aps,
pre,
nofootinbib]{revtex4-2}

\usepackage[T1]{fontenc}
\usepackage{graphicx}% Include figure files
\usepackage{dcolumn}% Align table columns on decimal point
\usepackage{bm}% bold math

\usepackage{amsmath,scalerel}	% Advanced maths commands
\usepackage{amssymb}	% Extra maths symbols
\usepackage{txfonts}
\usepackage[bb=libus]{mathalpha}
\usepackage{bbold}
\usepackage{relsize}
\usepackage{mathtools}
\usepackage{bigints}
\usepackage{sidecap}
\usepackage{caption,subcaption}
\usepackage{hyperref}

\usepackage{floatrow}
\usepackage{soul}
\usepackage{comment}
\usepackage[normalem]{ulem}

\usepackage{macros_ub}
\graphicspath{{./}{figures/}}

\begin{document}

\preprint{APS/123-QED}

\title{Collisionless relaxation to equilibrium distributions in cold dark matter halos: origin of the Navarro-Frenk-White profile} % Force line breaks with \\

\author{Uddipan Banik}
\email{uddipan.banik@princeton.edu,\;uddipanbanik@ias.edu}
\affiliation{Department of Astrophysical Sciences, Princeton University, 112 Nassau Street, Princeton, NJ 08540, USA\\Institute for Advanced Study, Einstein Drive, Princeton, NJ 08540, USA\\Perimeter Institute for Theoretical Physics, 31 Caroline Street N., Waterloo, Ontario, N2L 2Y5, Canada}
 %\altaffiliation[Also at ]{}%Lines break automatically or can be forced with \\
\author{Amitava Bhattacharjee}%
\email{amitava@princeton.edu}
\affiliation{Department of Astrophysical Sciences, Princeton University, 112 Nassau Street, Princeton, NJ 08540, USA}



%\affiliation{
%\affiliation{
%\affiliation{%
% This line break forced with %\textbackslash\textbackslash
%}%


\date{\today}% It is always \today, today,

\begin{abstract}

Collisionless self-gravitating systems such as cold dark matter halos are known to harbor universal density profiles despite the intricate non-linear physics of hierarchical structure formation in the $\Lambda$CDM paradigm. The origin of such states has been a persistent mystery, particularly because the physics of collisionless relaxation has remained poorly understood. To solve this long-standing problem, we develop a self-consistent quasilinear theory in action-angle space for the collisionless relaxation of inhomogeneous, self-gravitating systems by perturbing the governing Vlasov-Poisson equations. We obtain a quasilinear diffusion equation that describes the secular evolution of the mean coarse-grained distribution function $f_0$ of accreted matter in the fluctuating force field of a spherical isotropic halo. The diffusion coefficient not only depends on the fluctuation power spectrum but also on the evolving potential of the system, which reflects the self-consistency of the problem. Diffusive heating in the pre-assembled halo develops an $r^{-\gamma}$ cusp ($r$ is the halocentric radius) in the density profile of the accreted material. Accretion and relaxation in this $r^{-\gamma}$ inner cusp develops an $r^{-\beta}$ outer fall-off with $\beta \approx 5 - 2\gamma$ in the quasi-steady state. Spherical collapse theory dictates that a quasi-steady outer halo must settle to $\beta \approx 3$ since then the mass enclosed within a radially moving shell barely changes with time. This implies that the quasi-steady $\gamma$ must be approximately $1$, which is possible in the quasilinear framework only if (i) the pre-assembled halo harbors an $r^{-\gamma_\rmP}$ profile with $\gamma_\rmP \gtrsim 0.5$, (ii) its fluctuations are sufficiently correlated in time (red noise), and (iii) the initial value of $\gamma$ is smaller than $1$, implying that the $r^{-1}$ cusp is a neutral equilibrium. Self-consistent quasilinear relaxation therefore establishes the Navarro-Frenk-White (NFW) profile. We demonstrate for the first time how this profile emerges as a quasi-steady state of collisionless relaxation.

%\begin{description}
%\end{description}
\end{abstract}

\maketitle

%\tableofcontents

\section{Introduction}

Collisionless systems governed by long-range interactions are known to harbor non-thermal (non-Maxwellian) distribution functions. The two-body relaxation timescale can be extremely long in collisionless self-gravitating systems such as galaxies and cold dark matter (CDM) halos. Therefore, such systems are not expected to thermalize within the lifetime of the universe. Yet it is known that collisionless self-gravitating systems relax to steady state configurations often characterized by distribution functions (DFs) that are power law in energy, with a preference for particular power-law exponents. This is an outcome of collisionless relaxation that often occurs rapidly over a dynamical time, in which case it is referred to as violent relaxation \citep[][]{LyndenBell.67}, but sometimes as a secular process over several dynamical timescales. Despite several attempts over the last few decades, the origin of these states has remained a mystery.

Collisionless self-gravitating systems are described by the coupled, non-linear Vlasov-Poisson equations in a manner analogous to collisionless electrostatic plasmas. The Vlasov equation describes the evolution of the fine-grained DF $f$ under the action of the gravitational force, which is itself sourced by the density (zeroth velocity moment of the DF) through the Poisson equation. It is well known that the Vlasov equation admits a denumerably infinite set of Casimir invariants, of which the Boltzmann H-function (negative of the Boltzmann-Shannon entropy) is but one, and any positive definite function of the conserved quantities of the system is a valid steady state solution to the Vlasov equation (strong Jeans theorem). Why then do collisionless systems relax towards particular steady states? The answer lies in coarse-graining. The Vlasov equation evolves the fine-grained DF $f$. In reality, however, we can only measure the coarse-grained DF $f_0 = \left<f\right>$, obtained by some kind of averaging of the fine-grained DF, be it in actual observations, which are limited by instrumental resolution, or in numerical experiments, which are limited by grid resolution. The Vlasov equation predicts extreme filamentation of the fine-grained DF with small-scale structures all the way down to the free-streaming scale. The coarse-grained DF does not follow the Vlasov equation but a modified kinetic equation with a collision operator that encompasses the physics of Vlasov turbulence and kinetic instabilities. It is this effective collision operator that captures the small-scale (also known as sub-grid) physics of collective, collisionless relaxation in phase space and picks out a particular functional form for the coarse-grained DF $f_0$ in the steady state. This effective description of collisionless relaxation is very much in the same spirit as the effective field theories of particle physics and large-scale structure/cosmology. The collision operator in the modified kinetic equation can be very different from the Boltzmann operator (for example, it can be of the Balescu-Lenard form).

The kinetic equation for the relaxation of the coarse-grained DF of a stochastically perturbed, collisionless self-gravitating system such as a galaxy or cold dark matter (CDM) halo can be obtained using quasilinear theory (QLT), which involves perturbing the Vlasov-Poisson equations up to second order, followed by coarse-graining of the DF, i.e., spatial averaging of the DF for homogeneous systems and orbit/phase averaging for inhomogeneous ones. The physical setup we are concerned with in this paper is a halo that is assembling by (i) the accretion of matter into a pre-existing halo and (ii) the diffusive heating of the newly accreted matter by the stochastic gravitational perturbations of the halo. This yields a diffusion equation for the evolution of the coarse-grained DF of the accreted matter. Such a quasilinear diffusion equation (QLDE for short) has been recently derived in the context of collisionless plasmas by \citet[][]{Banik.etal.24} and \citet[][]{Banik.Bhattacharjee.24a} (see also references therein), and for collisionless systems governed by long-range interactions in general by \citet[][]{Chavanis.12,Chavanis.22,Chavanis.23}, who refers to it as the secular dressed diffusion equation\footnote{Such an equation describing the secular evolution of galaxies and halos was also anticipated by \citet[][]{Pontzen.Governato.13}.}.

In standard QLT, while the equation governing the time-evolution of the slowly evolving mean DF is exact, the fluctuations are assumed to obey linearized equations, when in reality, the fluctuations, too, obey nonlinear equations. As long as the perturbing forces are weaker than the mean gravitational force of the system, we are in the quasilinear regime. The long-time evolution of $f_0$ due to the interference of the linear perturbations is described the best by QLT if the quasilinear diffusion timescale is longer than the dynamical time of the system. In this paper we use QLT in the canonical action-angle variables \citep[c.f.][]{Hamilton.Fouvry.24} to study the evolution of the $f_0$ of an inhomogeneous CDM halo. In fusion plasma physics, a similar formulation of QLT using action-angle variables was pioneered by \citet[][]{Kaufman.72}. In the galactic context, QLT has been applied to the secular evolution of disk galaxies \citep[][etc]{Carlberg.Sellwood.85,Griv.etal.94,Hamilton.etal.24}, albeit not self-consistently. In the cosmological context, QLT has been used by \citet[][]{Ma.Bertschinger.04} to develop a kinetic equation for the evolution of the DF of an {\it average} halo under cosmological fluctuations (e.g., due to substructure). In this paper, we perturb the Vlasov-Poisson equations to obtain a QLDE that describes the relaxation of the {\it angle-averaged} or {\it phase-averaged DF} $f_0$ of an {\it individual} inhomogeneous halo. The key ingredient of this diffusion equation is the diffusion tensor, which depends on the fluctuation power-spectrum as well as the self-consistently evolving potential of the system.

What does $f_0$ look like in the fully non-linear setup? We can grow some intuition from the cosmological $N$-body simulations of a $\Lambda$CDM universe. It is difficult to measure $f_0$ precisely from these simulations due to the noise introduced by a finite number of simulation particles, but it is possible to measure its velocity moments, e.g., the density profile of a halo, which is the zeroth velocity moment of $f_0$ and is a smoother function. Early cosmological $N$-body simulations show a remarkable universality in the density profiles of CDM halos. Foremost among them is the Navarro-Frenk-White (NFW) profile \citep{Navarro.etal.97}, 
\begin{align}
\rho(r) = \dfrac{\rho_\rmc}{\dfrac{r}{r_\rms}{\left(1+\dfrac{r}{r_\rms}\right)}^2},
\label{rho_nfw}
\end{align}
with $r_\rms$ the scale radius and $\rho_\rmc$ a characteristic density, which is an excellent fit to the halo density, irrespective of the halo mass and concentration, power-law index of the initial power spectrum and cosmology. Later simulations, however, predict more diversity in the halo profiles. \citet[][]{Moore.etal.98b} finds that the inner halo harbors an $\sim r^{-1.4}$ cusp, much steeper than the NFW $r^{-1}$ cusp. \citet[][]{Navarro.etal.04}, on the other hand, find that most halos show an inner $r^{-1}$ cusp. Contrary to these results, high-resolution Aquarius \citep[][]{Navarro.etal.10} and Via Lactea II \citep[][]{Diemand.etal.07a} simulations find that the inner log-slope of the density profile becomes progressively shallower than $-1$ towards the center, akin to the \citet[][]{Einasto.65} profile. More recently, very high-resolution (zoom-in) cosmological simulations \citep[][]{Delos.White.23} have found the first halos to harbor steep $r^{-1.5}$ cusps akin to the \citet[][]{Moore.etal.98b} profile, which \citet[][]{Delos.White.23} refers to as prompt cusps. They point out that many of the halos eventually develop Einasto or NFW-like profiles around the prompt cusps as they grow in mass. All in all, there seem to exist particular preferred states in the landscape of halo profiles in $N$-body simulations.

We address the question of universality of halo profiles by answering the following key question: how does a halo assemble and relax, and what are the accessible relaxed states? We use the QLDE to model the collisionless relaxation of an inhomogeneous halo that is accreting and virializing, and find that in this process the halo relaxes to an NFW-like quasi-steady state under certain conditions. \citet[][]{Weinberg.01a,Weinberg.01b} also solves the QLDE, albeit for a different setup of a halo perturbed by orbiting subhalos/satellites, and for a limited range of initial halo profiles without a central cusp. He infers that weakly damped dipole modes excited by the orbiting satellites drive the secular relaxation of the halo towards an Einasto-like profile.

Our approach towards explaining the origin of halo profiles, while similar to that of \citet[][]{Weinberg.01a,Weinberg.01b}, is significantly different from most other previous work. We develop an Eulerian framework for the self-consistent evolution of the coarse-grained DF (under the quasilinear approximation), while previous literature has mainly focused on a Lagrangian framework for the orbital evolution of individual particles in a time-varying potential with the assumption of self-similarity and approximations for the orbital configuration. The secondary infall model of \citep[][]{Fillmore.Goldreich.84} and \citep[][]{Bertschinger.85} consists of a spherically symmetric self-similar solution for purely radial orbits that predicts an initial halo profile $\rho_i(r)\sim r^{-\gamma_\rmi}$ relaxing to a final halo profile $\rho(r)\sim r^{-\gamma_\rmf}$ with $\gamma_\rmf = 2$ for $\gamma_i \leq 2$ and $\gamma_\rmf = 3\gamma_\rmi/(1+\gamma_\rmi)$ for $\gamma_\rmi > 2$. It is, however, well known that a collisionless system with purely radial orbits is unstable to the formation of non-axisymmetric dipole and quadrupole (bar) modes \citep[][]{Antonov.73}. \citep[][]{Nusser.01} find that for non-zero but constant specific angular momentum per particle, one can obtain the \citet[][]{Fillmore.Goldreich.84} and \citet[][]{Bertschinger.85} slope of $\gamma_\rmf = 3\gamma_\rmi/(1+\gamma_\rmi)$ for all $\gamma_i$. The steep slope of $\gamma_\rmf = 2$ for $\gamma_\rmi < 2$ is eliminated due to the centrifugal barrier and non-zero periapse of particles moving along non-radial orbits. Interestingly, \citet[][]{Lu.etal.06} finds using 1D simulations that imposing isotropization of the particle velocities during collapse results in $\gamma_\rmf = 1$ for $\gamma_\rmi \leq 0.5$, which they interpret as a hint that the $r^{-1}$ NFW cusp may originate from orbit isotropization through violent relaxation. \citet[][]{Subramanian.00} and \citet[][]{Subramanian.etal.00} infer, using an effective fluid model for self-similar collapse based on the spherical Jeans equation, that the tangential velocity dispersion must be comparable to the radial dispersion (the halo must be sufficiently close to isotropy) in order to have $1\lesssim \gamma_\rmf < 2$. \citet[][]{Dehnen.McLaughlin.05} finds, using the spherical Jeans equation, that as long as $\rho/\sigma^3_r$ ($\sigma_r$ is the radial velocity dispersion) is a power-law in $r$, the density has an inner log-slope $\gamma \approx 0.81$ and an outer log-slope $\beta \approx 3.44$, with some dependence on the anisotropy parameter. \citet[][]{Dekel.etal.03a,Dekel.etal.03b} finds that halos tend to relax towards a central cusp with $\gamma_\rmf$ slightly larger than $1$. They argue that cored halos with $\gamma_\rmf < 1$ exert compressive tidal forces on the infalling subhalos, which therefore survive disruption and inspiral all the way to the center under dynamical friction, resulting in $\gamma_\rmf \gtrsim 1$. This, however, does not take into account core-stalling \citep[][]{Banik.vdBosch.21a,Banik.vdBosch.22}, the stalling of subhalo inspiral due to vanishing dynamical friction in cored galaxies, found in later high resolution idealized simulations \citep[][]{Read.etal.06c,Goerdt.etal.10,Inoue.11,Cole.etal.12}. \citet[][]{Dalal.etal.10} finds that a self-similar solution with adiabatic invariance of the radial action yields a halo profile with a central core and a gradual Einasto-like roll-over of the log-slope akin to the profiles obtained from the high resolution Via Lactea II and Aquarius simulations. \citet[][]{Pontzen.Governato.13} uses a prescription for constrained entropy maximization to obtain the distributions of angular momenta and radial actions that match well with those of halos in the GHALO simulation. To our knowledge, though, no such entropy-based argument has ever correctly reproduced the halo density profile.

Whether CDM halos possess a universal profile at all, and whether it is NFW-like, Einasto-like or prompt cusp-like or something else altogether, has been a matter of long-drawn controversy. This is mainly because the physics of collective, collisionless relaxation has remained poorly understood. We adopt an alternate route, fundamentally different from the above approaches but in the same spirit as \citep[][]{Weinberg.01a} and \citep[][]{Weinberg.01b}. Instead of looking at the evolution of the halo density profile directly, we build a QLT for the collisionless relaxation of the mean coarse-grained DF $f_0$ from first principles (Vlasov-Poisson equations), formulate a governing diffusion equation for $f_0$, look for its steady state solutions as well as their accessibility, and identify the corresponding halo density profiles. We find that the NFW profile emerges from this process of collisionless relaxation, provided that the pre-assembled halo harbors a sufficiently steep profile, its noise is sufficiently red, and the initial profile of the accreted matter is sufficiently shallow. A key aspect in which our work differs from those of \citep[][etc]{Fillmore.Goldreich.84,Bertschinger.85,Nusser.01} is that our formulation enables the identification of more than a single power-law. Moreover, unlike these studies, we do not make any specific assumptions about the orbital configuration, but assume velocity isotropy for $f_0$, which appears to be an essential feature of a virialized halo, especially in the inner region. Our formulation is based on the Vlasov-Poisson equations and therefore has a wider scope than the effective fluid model of \citep[][etc]{Subramanian.00,Subramanian.etal.00,Dehnen.McLaughlin.05}.

This paper is organized as follows. Section~\ref{sec:resp_theory} introduces the perturbative (linear and quasilinear) response theory for the relaxation of perturbed collisionless self-gravitating systems governed by the Vlasov-Poisson equations. In Section~\ref{sec:quasilinear}, we use this theory to study the assembly and relaxation of a spherical isotropic halo. We derive the QLDE that describes the evolution of the mean coarse-grained DF of matter accreted in a stochastically fluctuating halo, which we solve to obtain the quasi-steady $f_0$ and the corresponding halo profile. We summarize our findings in section~\ref{sec:discussion_summary}.

\section{Response theory for collisionless self-gravitating systems}\label{sec:resp_theory}

%\begin{figure}
%\centering
%\includegraphics[width=1\textwidth]{illustration.jpeg}
%\caption{Secular evolution of the halo due to the orbital inspiral of the subhalos under dynamical friction and the associated diffusive heating of the halo.}
%\label{fig:illustration}
%\end{figure}

\subsection{Physical setup}

We study the evolution of a self-gravitating system such as a galaxy or dark matter halo by tracking how its different parts gravitationally interact with each other. We formulate a theory for the response of a system to a perturbing potential $\Phi_\rmP$. The response can be modeled as a linear perturbation if the perturbing force is weaker than the mean gravitational force of the system itself. In this paper, we develop a working model for how a halo assembles over time. Consider a spherically symmetric halo that is fluctuating (virializing). As the halo gravitationally accretes new matter, it gets heated by the fluctuating halo and relaxes to a quasi-steady distribution different from the initial one. As more matter gets accreted, it experiences stochastic heating by the modified halo. This is how the halo grows and relaxes.

During this process of stochastic heating, energy gets transferred from the perturber (fluctuating halo) to the system (accreted material) in a diffusive manner. This is because the DF $f_0$ of the system is typically a monotonically decreasing function of energy, so that there exist more particles with lower energy than the perturber, than with higher energy. As a result, more particles gain energy from than lose energy to the perturber. Since the total energy of the system and the perturber is conserved, the perturber cools and experiences dynamical friction \citep[][]{Chandrasekhar.43,Tremaine.Weinberg.84} and the system heats up. In this paper, we focus on the relaxation of the system and not on that of the perturber. As alluded to above, we are interested in the scenario where a system of accreted matter is heated by the gravitational fluctuations in the pre-assembled halo which acts as the perturber.

We set up the problem in the following way. We compute the linear response of the system to the perturber using the linearized Vlasov-Poisson equations for the system. This response is collectively dressed by the mutual self-gravity of the particles. The perturber locally enhances the halo density, which gets amplified due to self-gravity. The particle distribution not only gets denser but is also heated in the process. This heating manifests as an increase in the velocity dispersion of the system and is described by a quasilinear (second order) response theory, which yields a quasilinear diffusion equation (QLDE) for the diffusive broadening of the mean DF $f_0$ of the system. As the system heats up, the change in $f_0$ changes its density profile and consequently its potential through the Poisson equation, which in turn changes the diffusion coefficient. This self-consistent evolution is a crucial ingredient of our theory.

\subsection{Governing equations}

Now we mathematically formulate the theory for collisionless relaxation. A self-gravitating system is characterized by the DF or phase space ($\bx,\bv$) density of its constituent particles, $f(\bx,\bv,t)$. The general equations governing the relaxation of a collisionless self-gravitating fluid such as a CDM halo or a galaxy are the collisionless Boltzmann or Vlasov and Poisson equations,

\begin{align}
&\frac{\partial f}{\partial t} + \left[f,H\right] = 0,\nonumber\\
&\nabla^2 \Phi = 4\pi G \int d^3v\, f,
\end{align}
where $H=v^2/2 + \Phi$ denotes the Hamiltonian, with $\Phi$ the gravitational potential, and $\left[f,H\right]$ denotes the Poisson bracket. We describe the inhomogeneous galaxy or halo in terms of the canonical angle-action variables, $\left(\bw,\bI\right)$, with $\bw = \left(w_r,w_\theta,w_\phi\right)$ and $\bI = \left(I_r,L,L_z\right)$. Here, $I_r$ is the radial action, $L$ is the angular momentum, and $L_z$ is its $z$ component, while $w_r$ is the radial angle, $w_\theta$ is the angle in the orbital plane and $w_\phi$ is the longitude of the ascending node that is constant for a spherically symmetric halo. The Poisson bracket is given by $\left[f,H\right]=\nabla_{\bw} f \cdot \nabla_{\bI} H - \nabla_{\bI} f \cdot \nabla_{\bw} H$.

The Hamiltonian of the system, perturbed by an external perturbing potential $\Phi_\rmP$, can be written as $H = H_0 + \Phi_\rmP + \Phi'$ with $H_0 = v^2/2 + \Phi_0$, $\Phi_0$ the quasi-equilibrium halo potential and $\Phi'$ the self-consistent potential sourced by the perturber-induced response through the Poisson equation. We consider $\Phi_\rmP$ to be a stochastic potential sourced by inhomogeneities in the perturber. The Vlasov equation is difficult to solve in its full generality due to the non-linearity in both $\bw$ and $\bI$, and hence, one must resort to perturbation theory to make analytical progress. If the strength of the perturber potential, $\Phi_\rmP$, is smaller than $\sigma^2_0$, where $\sigma_0$ is the velocity dispersion of the unperturbed quasi-equilibrium system, then the perturbation in $f$ can be expanded as a power series in the perturbation parameter, $\epsilon \sim \left|\Phi_\rmP\right|/\sigma^2_0$, i.e., $f = f_0 + \epsilon f_1 + \epsilon^2 f_2 + ...\,$; the self-consistent potential $\Phi'$ can also be expanded accordingly as $\Phi' = \epsilon \Phi_1 + \epsilon^2 \Phi_2 + ...\,$.

\subsection{Equilibrium profile}\label{sec:eqbm}

Before discussing the perturbative response theory for collisionless relaxation, let us describe the equilibrium model for the system. We assume the quasi-equilibrium density profile and potential of the system to be spherically symmetric. Later in the paper, we would require the functional dependencies of the energy $E$, radial action $I_r$, angular momentum $L$, frequencies ${\bf \Omega} = \partial E/\partial \bI$ and the quasi-equilibrium DF $f_0$ on the semi-major axis length $a$ of an orbit, which we state as follows.

The equilibrium density $\rho_0(r)$ of the system is related to its equilibrium potential $\Phi_0(r)$ through the spherically symmetric Poisson equation:

\begin{align}
\frac{1}{r^2}\frac{\rmd}{\rmd r}\left(r^2\frac{\rmd \Phi_0}{\rmd r}\right) = 4\pi G \rho_0(r).
\label{Poisson_eq_rad}
\end{align}
It can be easily seen that if $\rho_0(r)\sim r^{-\gamma}$ with $0<\gamma<2$ (in the inner halo, for $r<r_\rms$), the corresponding potential is $\Phi_0(r) = \Phi_\rmc\left(1-{\left(r/r_\rms\right)}^{2-\gamma}\right)$ with $\Phi_\rmc = -G M_0/\left(2-\gamma\right)$ the central potential, $M_0$ the system mass and $r_\rms$ the scale radius. We only consider $\gamma<2$, so that both the enclosed mass $M_0(r)=4\pi\int_0^r \rmd r'\,r'^2\rho_0(r')$ and the potential are finite at $r\to 0$. The energy $E$ scales as $\sim \Phi_\rmc\left[1-{\left(a/r_\rms\right)}^{2-\gamma}\left(\left({\left(1+e\right)}^{4-\gamma}-{\left(1-e\right)}^{4-\gamma}\right)/4e\right)\right]$ with $a=\left(r_\rma+r_\rmp\right)/2$ the length of the semi-major axis, $e = \left(r_\rma-r_\rmp\right)/\left(r_\rma+r_\rmp\right)$ the eccentricity and $r_\rma$ and $r_\rmp$ the apo- and peri-centric radii that satisfy $E = \Phi_0(r) + L^2/2 r^2$. The angular momentum $L$ is given by $L^2 = L^2_\rmc \left[{\left(1-e^2\right)}^2/\left(2\left(2-\gamma\right)e\right)\right]\left[{\left(1+e\right)}^{2-\gamma} - {\left(1-e\right)}^{2-\gamma}\right]$ with $L_\rmc\sim \sqrt{G M_0 r_\rms}\,{\left(a/r_\rms\right)}^{2-\gamma/2}$ the circular angular momentum. The radial action also scales as $\sim a^{2-\gamma/2}$. Both the tangential frequency $\Omega_\theta$ (in the orbital plane) and the radial frequency $\Omega_r$ scale as $\sim a^{-\gamma/2}$, with weak dependence on $e$. The equilibrium DF can be obtained by Eddington inversion of the density profile \citep[][]{Binney.Tremaine.08}, and can be shown to scale as follows for $0 < \gamma < 2$ \citep[][]{Dehnen.93}:

\begin{align}
f_0\left(E\right) \sim {\left(\Psi_\rmc - \calE\right)}^{-{\textstyle{\frac{6-\gamma}{2\left(2-\gamma\right)}}}} \sim a^{{\textstyle{\frac{\gamma}{2}}}-3},
\end{align}
with $\calE = -E$ and $\Psi_\rmc = -\Phi_\rmc$. 

If $\rho_0(r)\sim r^{-\beta}$ with $\beta > 3$ (in the outer halo, for $r>r_\rms)$, which is necessarily the case so that the total mass is finite, the potential $\Phi_0(r)$ scales as $-G M_0/r$ and the energy as $\sim -G M_0/2a$. The frequencies scale as $\sim a^{-3/2}$ and the angular momentum and radial action as $\sim a^{1/2}$. The DF, obtained by Eddington inversion, scales as

\begin{align}
f_0\left(E\right) \sim {\calE}^{\,\beta-3/2} \sim a^{{\textstyle{\frac{3}{2}}}-\beta},
\end{align}
For $\beta = 3$, the various quantities scale similarly as above except for logarithmic corrections in $a$.

\subsection{Linear response theory}\label{sec:lin_resp_theory}

The first-order response of the system is described by the linearized Vlasov-Poisson equations,

\begin{align}
&\frac{\partial f_1}{\partial t} + \left[f_1,H_0\right] + \left[f_0, \Phi_\rmP\right] + \left[f_0, \Phi_1\right] = 0,\nonumber\\
&\nabla^2 \Phi_1 = 4\pi G \int d^3v\, f_1.
\label{lin_CBE_Poisson}
\end{align}
We assume that the unperturbed, quasi-equilibrium $f_0$ is phase/angle-averaged and is therefore only a function of the actions (strong Jeans theorem). Expanding the linear perturbations as Fourier series in angles and performing the Laplace transform in time, we obtain the following expression for the Fourier-Laplace transform of the linear response, $\Tilde{f}_{1\boldell}(\bI,\omega)$, in terms of that of the perturber potential, $\Tilde{\Phi}_{\rmP\boldell}(\bI,\omega)$ and the self-consistent potential, $\Tilde{\Phi}_{1\boldell}(\bI,\omega)$ (see Appendix~\ref{App:lin_resp} for a detailed derivation):

\begin{align}
\Tilde{f}_{1\boldell}(\bI,\omega) = - \,\boldell \cdot \frac{\partial f_0}{\partial \bI} \,\frac{\Tilde{\Phi}_{\rmP\boldell}(\bI,\omega) + \Tilde{\Phi}_{1\boldell}(\bI,\omega)}{\omega - \boldell \cdot {\bf \Omega}},
\label{f1l}
\end{align}
where tilde indicates the Laplace transform. Here, ${\bf\Omega} = \nabla_{\bI}H_0=\left(\Omega_r,\Omega_\theta,\Omega_\phi\right)$ denote the unperturbed orbital frequencies of the particles ($\Omega_\phi=0$ for a spherically symmetric system since $H_0$ is independent of $L_z$ and the longitude of ascending node is a constant). We have assumed the initial perturbation $f_{1\boldell}\left(\bI,t=0\right)=0$.

When the self-consistent potential $\Phi_1$ is comparable to the perturber potential $\Phi_\rmP$, we must include the coupling of $\Phi_1$ to the density perturbation $\rho_1 = \int \rmd^3 v\, f_1$ through the Poisson equation. This requires us to expand the Fourier-Laplace coefficients in terms of bi-orthogonal basis functions as outlined in Appendix~\ref{App:lin_resp}, which yields the following response equation:

\begin{align}
\Tilde{\mathbb{\ba}}(\omega) = {\left(\varmathbb{I}-\varmathbb{M}(\omega)\right)}^{-1} \varmathbb{M}(\omega)\, \Tilde{\mathbb{\bb}}(\omega).
\label{lin_resp_eq}
\end{align}
Here $\varmathbb{I}$ denotes the identity matrix, and $\varmathbb{M}$ indicates the response matrix given by

\begin{align}
\varmathbb{M}_{pq} (\omega) = \frac{{\left(2\pi\right)}^3}{4\pi G} \sum_{\boldell} \int \rmd \bI\; \boldell \cdot \frac{\partial f_0}{\partial \bI}\, \frac{\psi^{(p)\ast}_{\boldell}(\bI) \psi^{(q)}_{\boldell}(\bI)}{\omega - \boldell\cdot \Omega}\,.
\label{resp_matrix}
\end{align}
The matrix, $\left(\varmathbb{I}-\varmathbb{M}\right)$, denotes the dielectric tensor. $\psi^{(p)}_{\boldell}\left(\bI\right)$ denotes the Fourier coefficient (of the $\boldell$ mode) with respect to the angles of the basis function $\psi^{(p)}(\bx)$. The potentials are expanded in terms of these basis functions as $\Phi_1(\bx,t) = \sum_{p} a_p(t) \psi^{(p)} (\bx)$ and $\Phi_\rmP(\bx,t) = \sum_{p} b_p(t) \psi^{(p)} (\bx)$. $\Tilde{\mathbb{\ba}}$ ($\Tilde{\mathbb{\bb}}$) denotes the Laplace transform of $\mathbb{\ba}$ ($\mathbb{\bb}$). Equation~(\ref{lin_resp_eq}) manifests the dressing of the response due to self-gravity (akin to dielectric polarization in a plasma). The response matrix, which would be zero in the absence of self-gravity, encodes all information about this dressing. The halo particles gravitationally interact with each other, which causes them to experience the dressed and not the bare potential of the perturber. Performing the inverse Laplace transform of the response equation~(\ref{lin_resp_eq}) shows that the temporal response of the $\boldell$ mode consists of three terms: a continuum response that evolves as $\exp{\left[-i\boldell\cdot{\bf \Omega}t\right]}$ and denotes the oscillations of the response at the unperturbed orbital frequencies (which eventually phase-mixes away), a forced response or wake that follows the temporal dependence of the perturber (responsible for dynamical friction \citep[][]{Tremaine.Weinberg.84,Weinberg.89,Banik.vdBosch.21a,Kaur.Sridhar.18,Kaur.Stone.22}) and a set of coherent oscillations or discrete Landau/point modes oscillating at frequencies $\omega_n$ that follow the dispersion relation, ${\rm det}\left(\varmathbb{I}-\varmathbb{M}\left(\omega_n\right)\right)=0$ (see Appendix~\ref{App:lin_resp} for a detailed derivation of the temporal linear response).

Self-gravity significantly amplifies the response when the perturber is near-resonant with the particles ($\omega \sim \boldell\cdot{\bf\Omega}$). Faster perturbation ($\omega \gtrsim \boldell\cdot{\bf\Omega}$) is nearly unaffected by collective dressing, in which case the response matrix $\varmathbb{M}\approx 0$ and the dielectric tensor $\varmathbb{I}-\varmathbb{M} \approx \varmathbb{I}$, $\Phi_1$ may be neglected relative to $\Phi_\rmP$ \citep[][]{Fouvry.etal.21}, and we have a simpler expression for the linear response:

\begin{align}
\Tilde{f}_{1\boldell}(\bI,\omega) = - \,\boldell \cdot \frac{\partial f_0}{\partial \bI} \,\frac{\Tilde{\Phi}_{\rmP\boldell}(\bI,\omega)}{\omega - \boldell \cdot {\bf \Omega}}.
\label{f1l_non_sg}
\end{align}
In the case of slower perturbations ($\omega \lesssim \boldell\cdot{\bf\Omega}$), the determinant of the large-scale (small $p$ and $q$) part of the dielectric tensor is less than unity but nearly independent of $\omega$, while that of the small-scale (large $p$ and $q$) part is close to unity. Therefore, self-gravity only enhances the response when (i) the perturber is near-resonant with or slower than the halo particles and (ii) the perturber acts on scales larger than the scale radius of the system. 

\subsection{Second-order response theory}\label{sec:2nd_resp_theory}

The linear perturbations $f_1$ and $\Phi_\rmP + \Phi_1$ non-linearly couple and drive the evolution of $f$ at second order. Physically, the linear response $f_1$ describes the density enhancement around the perturber, while the second order response $f_2$ describes the enhancement of velocity dispersion. The second-order response is described by the following evolution equations for $f_2$ and $\Phi_2$:

\begin{align}
&\frac{\partial f_2}{\partial t} + \left[f_2,H_0\right] + \left[f_1, \Phi_\rmP\right] + \left[f_1, \Phi_1\right] + \left[f_0,\Phi_2\right] = 0,\nonumber\\
&\nabla^2 \Phi_2 = 4\pi G \int d^3v\, f_2.
\label{2nd_CBE_Poisson}
\end{align}
The evolution of $f_2$ is guided by that of the linear fluctuations, $f_1$ and $\Phi_1$, which we have already computed using linear response theory. 

As before, we can solve the above equations in the Fourier space of angles. The evolution of the mean background DF, averaged over the angles and the random phases of the linear fluctuations, $f_0 = \int \rmd^3 w\, f/{\left(2\pi\right)}^3 \approx f_{1\boldell=0} + f_{2\boldell=0} = f_{2\boldell=0}$ (note that $f_{1\boldell=0} = 0$ from equation~[\ref{f1l}]), can be studied by taking the $\boldell \to 0$ limit of the second order response, $f_{2\boldell}$. This yields (see Appendix~\ref{App:QL_resp} for details)

\begin{align}
\frac{\partial f_{0}}{\partial t} = i\sum_{\boldell}\boldell\cdot\frac{\partial}{\partial \bI}\left<f^\ast_{1\boldell}\left(\bI,t\right)\Phi_{\boldell}\left(\bI,t\right)\right>,
\label{QL_eq}
\end{align}
where we have absorbed the factor $\epsilon^2$ in the correlation of $f_{1\boldell}^\ast$ and $\Phi_{\boldell}$ in the RHS. $f_{1\boldell}$ is the Fourier coefficient of $f_1$, while $\Phi_{\boldell}$ is equal to $\Phi_{\rmP\boldell} + \Phi_{1\boldell}$, $\Phi_{\rmP\boldell}$ and $\Phi_{1\boldell}$ being the Fourier coefficients of $\Phi_\rmP$ and $\Phi_1$ respectively. The brackets denote an ensemble average over the random phases of the fluctuations\footnote{Under the ergodic hypothesis, this is the same as a temporal average with a window that is equal to at least the correlation time of the fluctuations.}. The unperturbed mean DF $f_0$ is not a stationary quantity, rather it evolves secularly on a timescale longer than the mean dynamical time via the above quasilinear equation. Upon substituting the expressions for $f_{1\boldell}$ and $\Phi_{\boldell}$ obtained using linear response theory in the above equation, and taking the long time limit such that the Landau modes have damped away (assuming there are no instabilities), we obtain the following form for the quasilinear diffusion equation or QLDE (see Appendix~\ref{App:QL_resp} for a detailed derivation):

\begin{align}
\frac{\partial f_0}{\partial t} = \sum_{\boldell}\boldell\cdot\frac{\partial}{\partial \bI} \left(D_{\boldell}\left(\bI\right) \, \boldell\cdot\frac{\partial f_0}{\partial \bI}\right),
\label{QL_eq1}
\end{align}
with the diffusion coefficient $D_{\boldell}\left(\bI\right)$ given by

\begin{align}
D_{\boldell}\left(\bI\right) = {\left|{\left(\varmathbb{I}-\varmathbb{M}\left(\boldell\cdot\bf{\Omega}\right)\right)}^{-1}_{pq} B_q \psi_{\boldell}^{(p)}\left(\bI\right)\right|}^2 \calC_{\omega}\left(\boldell\cdot{\bf\Omega}\right),
\label{Dl}
\end{align}
where the Einstein summation convention is implied and $\varmathbb{M}_{pq}\left(\boldell\cdot\bf{\Omega}\right)$ is given by

\begin{align}
&\varmathbb{M}_{pq} \left(\boldell\cdot\bf{\Omega}\right) = \frac{{\left(2\pi\right)}^3}{4\pi G} \sum_{\boldell'} \int \rmd \bI' \frac{\partial f_0}{\partial E'}\, \psi^{(p)\ast}_{\boldell'}(\bI') \psi^{(q)}_{\boldell'}(\bI') \nonumber\\
&\times \left[{\left(\frac{\boldell\cdot{\bf \Omega}}{\boldell'\cdot{\bf \Omega'}}-1\right)}^{-1} - i\pi\,\boldell'\cdot{\bf \Omega'}\delta\left(\boldell\cdot{\bf\Omega} - \boldell'\cdot{\bf\Omega'}\right)\right].
\label{resp_matrix_Omega}
\end{align}
Here we have split the response matrix into the non-resonant $\left(\boldell\cdot{\bf\Omega} \neq \boldell'\cdot{\bf\Omega'}\right)$ principal value part and the resonant $\left(\boldell\cdot{\bf\Omega} = \boldell'\cdot{\bf\Omega'}\right)$ part. In deriving the above diffusion equation, we have assumed the perturber potential to be a generic red noise:

\begin{align}
\left<b^\ast_q\left(t\right) b_{q'}\left(t'\right)\right> = B^\ast_q B_{q'} \calC_t\left(t-t'\right),
\label{subhalo_model}
\end{align}
with $\calC_t$ the temporal correlation function that is equal to $\delta\left(t-t'\right)$ for white/uncorrelated noise. The Fourier transform of the correlation function is given by $\calC_{\omega}$, which, for white noise, is simply equal to $1$. Note that the diffusion coefficient consists of three key ingredients: (i) the spatial power spectrum of the perturbations, (ii) the temporal power spectrum and (iii) the collective dressing of the perturbations, denoted by the dielectric tensor, $\varmathbb{I}-\varmathbb{M}$. We have assumed that all Landau modes have damped away, i.e., we are looking at the long time relaxation of the system at $t\gtrsim 1/\gamma_0$, where $\gamma_0$ is the damping rate of the least damped Landau mode. Under these assumptions, we find that $f_0$ evolves under the above QLDE, also known as the secular dressed diffusion equation \citep[][]{Chavanis.12,Chavanis.22,Chavanis.23}.

If the perturber acts on scales larger than the semi-major axis $a(\bI)$ of the orbit under consideration, then $\boldell\cdot{\bf\Omega}\gtrsim \boldell'\cdot{\bf\Omega'}$ for the majority of $\bI'$ in the integrand of $\varmathbb{M}_{pq}\left(\boldell\cdot{\bf\Omega}\right)$ (equation~[\ref{resp_matrix_Omega}]), which implies that $\varmathbb{M}_{pq}\left(\boldell\cdot{\bf\Omega}\right)\approx 0$. In other words, self-gravity may be neglected for rapidly orbiting particles confined well within the perturbing potential \cite[][]{Fouvry.etal.21}. This enables a substantial simplification of the QLDE. Modeling the fluctuating perturber as

\begin{align}
\left<\Phi^\ast_{\rmP\boldell}\left(\bI,t\right) \Phi_{\rmP\boldell}\left(\bI,t'\right)\right> = {\left|\Psi_{\rmP\boldell}\left(\bI\right)\right|}^2\, \calC_t\left(t-t'\right),
\label{subhalo_model_non_sg}
\end{align}
where $\Psi_{\rmP\boldell}\left(\bI\right)$ denotes the Fourier transform of the spatial part, the diffusion coefficient can be simplified into

\begin{align}
D_{\boldell}\left(\bI\right) = {\left|\Psi_{\rmP\boldell}\left(\bI\right)\right|}^2 \calC_{\omega}\left(\boldell\cdot{\bf\Omega}\right).
\label{Dl_non_sg}
\end{align}
Note that it has the units of potential squared times time.

The QLDE describes how the smooth distribution of the system heats up under stochastic gravitational perturbations. Of course, this assumes that the force perturbations are weaker than the mean gravitational force. It should be borne in mind that the QLDE provides a good description of the long-term relaxation of the system over several dynamical times but not of its violent relaxation over a few. 

\section{Quasilinear theory for collisionless relaxation}\label{sec:quasilinear}

\subsection{Quasilinear diffusion equation}

Now we study the collisionless relaxation of the system by evolving the phase-averaged DF $f_0$ via the quasilinear equation~(\ref{QL_eq1}), which can be recast into the following form:

\begin{align}
\frac{\partial f_0}{\partial t} = \frac{\partial}{\partial I_i} \left(D_{ij}\left(\bI\right)\frac{\partial f_0}{\partial I_j}\right),
\label{QL_diff_eq_gen}
\end{align}
with the diffusion tensor $D_{ij}$ given by

\begin{align}
D_{ij}\left(\bI\right) = \sum_{\boldell} \ell_i \ell_j\, {\left|{\left(\varmathbb{I}-\varmathbb{M}\left(\boldell\cdot\bf{\Omega}\right)\right)}^{-1}_{pq} B_q \psi_{\boldell}^{(p)}\left(\bI\right)\right|}^2 \calC_{\omega}\left(\boldell\cdot{\bf\Omega}\right)
\label{diff_coeff_QL}
\end{align}
in general, and by

\begin{align}
D_{ij}\left(\bI\right) = \sum_{\boldell} \ell_i \ell_j\, {\left|\Psi_{\rmP\boldell}\left(\bI\right)\right|}^2\, \calC_{\omega}\left(\boldell\cdot{\bf\Omega}\right),
\label{diff_coeff_QL_non_sg}
\end{align}
when collective dressing is inefficient. 

Let us now make a series of simplifying assumptions to make the QLDE analytically tractable and glean the essential physics of collisionless relaxation. First, we assume that the system is spherically symmetric and isotropic in velocities. In this case $f_0$ can be described as a function of the energy $E$, i.e., $f_0$ is an ergodic distribution $f_0(E)$ \citep[][]{Binney.Tremaine.08}. This enables us to rewrite $\boldell\cdot\partial f_0/\partial \bI$ as $\boldell\cdot{\bf \Omega}\, \partial f_0/\partial E$, which reduces the above QLDE into the following one dimensional diffusion equation in energy:

\begin{align}
\frac{\partial f_0}{\partial t} = \sum_{\boldell} \boldell\cdot{\bf \Omega} \frac{\partial}{\partial E} \left( \boldell\cdot{\bf \Omega}\, D_{\boldell}\left(\bI\right)\frac{\partial f_0}{\partial E}\right),
\label{QL_diff_eq_1D}
\end{align}
with $D_{\boldell}\left(\bI\right)$ given by equation~(\ref{Dl}). Here we have used the fact that ${\bf \Omega} = \partial H_0/\partial \bI = \partial E/\partial \bI$. Although ${\bf \Omega}$ and $D_{\boldell}$ depend on the angular momentum $L$, this dependence is much weaker than that on $E$ for a spherically symmetric and isotropic system. 

Next, we assume that the perturbing potential is also spherically symmetric. In this case, the orbital energies and radial actions (eccentricities) of the particles gradually increase, while their angular momenta are conserved. The QLDE can be recast into the following one dimensional diffusion equation in $I_r$:

\begin{align}
\frac{\partial f_0}{\partial t} = \frac{\partial}{\partial I_r}\left(D\left(L,I_r\right)\frac{\partial f_0}{\partial I_r}\right),
\label{QLDE_Ir}
\end{align}
where we have used the fact that $\Omega_r = \partial H_0/\partial I_r$. The diffusion coefficient $D\left(L,I_r\right)$ is given by

\begin{align}
D\left(L,I_r\right) = \sum_{\ell_r} \ell_r^2\, {\left|{\left(\varmathbb{I}-\varmathbb{M}\left(\ell_r\Omega_r\right)\right)}^{-1}_{pq} B_q \psi_{\ell_r}^{(p)}\left(L,I_r\right)\right|}^2 \calC_{\omega}\left(\ell_r\Omega_r\right),
\label{diff_coeff_QL_Ir}
\end{align}
which simplifies to 

\begin{align}
D\left(L,I_r\right) = \sum_{\ell_r} \ell_r^2\, {\left|\Psi_{\rmP\ell_r}\left(L,I_r\right)\right|}^2 \calC_{\omega}\left(\ell_r\Omega_r\right),
\label{diff_coeff_QL_Ir_non_sg}
\end{align}
when collective dressing is negligible. Note that the $\ell_\phi$ dependence has dropped out due to the assumption of a spherically symmetric perturber, in which case $\Psi_{\rmP\boldell} = \Psi_{\rmP\ell_r}\delta_{\ell_\phi,0}$. The diffusion coefficient depends on the actions mainly through the semi-major axis $a$, with mild dependence on the eccentricity $e$.

In the present scenario of the relaxation of accreted matter in a fluctuating halo, dressing does not introduce significant additional $\bI$ dependence to the diffusion coefficient. Therefore, to obtain essential scalings in this paper, we shall neglect self-gravity of the perturbations and work with the simpler version of the diffusion coefficient given in equation~(\ref{diff_coeff_QL_Ir_non_sg}). Even so, we have included self-gravity in the formal theory for the sake of completeness and applicability to scenarios where dressing plays an important role (e.g., in dynamically cold systems like galactic disks).

%Finally, we assume for our fiducial case that the perturber acts as a white noise, for which $\calC_t\left(t-t'\right)=\delta\left(t-t'\right)$ and $\calC_\omega\left(\ell_r\Omega_r\right) = 1$. In other words, any two subsequent perturbations are uncorrelated, or even if they are correlated, the correlation time is smaller than the radial orbital period, $2\pi/\boldell \cdot {\bf \Omega_r}$ at $\bI$. We explore the impact of red noise in section~\ref{sec:red_noise}. 

\subsection{Steady state solution}\label{sec:steady_state}

Before obtaining the time-dependent solution, let us explore the steady state solution to the QLDE (equation~[\ref{QLDE_Ir}]):

\begin{align}
{\rm Flux} = -D\left(L,I_r\right)\frac{\partial f_0}{\partial I_r} = {\rm constant}.
\label{steady_state_cond_const_flux}
\end{align}
Note that the diffusive flux is either positive or zero for a stable system since $\partial f_0/\partial I_r \leq 0$. This implies that such a system always tends to heat up under stochastic perturbations. If the flux is zero, then we have the trivial solution that $f_0$ is a constant. The corresponding $\rho_0$ and $\Phi_0$ can still be non-trivial functions of $r$, as we discuss in section~\ref{sec:zero_flux}. 

If the flux is a non-zero constant, then we have a non-trivial solution for $f_0(I_r)$ or $f_0(E)$. This of course depends on the $I_r$ dependence of the diffusion coefficient, which in turn depends on the spatiotemporal nature of the perturbing potential. For a spherically symmetric perturber, the spatial dependence is naturally of the following form:

\begin{align}
\Phi_\rmP(r) \sim
\begin{cases}
r^{2-\gamma_\rmP}/\left(2-\gamma_\rmP\right), \quad \gamma_\rmP < 3,\gamma_\rmP \neq 2,\nonumber\\
\ln{\left(r/r_\rms\right)},\qquad \quad \;\;\; \gamma_\rmP = 2,\nonumber\\
-r^{-1}, \qquad \qquad \;\;\;\,\, \gamma_\rmP > 3,
\end{cases}
\end{align}
where the density profile of the perturber, $\rho_\rmP(r)$ scales as $\sim r^{-\gamma_\rmP}$. This implies that $\Psi_{\rmP\ell_r}\sim a^{2-\gamma_\rmP}$ for $\gamma_\rmP<3$ and $\gamma_\rmP \neq 2$, $\ln{\left(a/r_\rms\right)}$ for $\gamma_\rmP = 2$ and $a^{-1}$ for $\gamma_\rmP > 3$, with a mild dependence on $e$ (for $\ell_r \neq 0$ modes that have a non-zero contribution to the diffusion coefficient). We assume that the perturbing mass is fluctuating in time as a generic red noise characterized by $\calC_{\omega}\left(\ell_r\Omega_r\right)$, which is equal to $1$ for $\ell_r\Omega_r t_\rmc \lesssim 1$ (white noise) and $\sim {\left(\ell_r\Omega_r t_\rmc\right)}^{-n}$ for $\ell_r\Omega_r t_\rmc \gtrsim 1$, with $t_\rmc$ the correlation time. 

Collective dressing does not introduce significant $a$ dependence to the diffusion coefficient since the response matrix is independent of $a$ in both small and large $a$ limits. Therefore, dressing may be neglected while deriving the $a$ (equivalently $I_r$ or $E$) scalings of the various quantities, in which case the diffusion coefficient $D(L,I_r)$ bears a much simpler expression given by equation~(\ref{diff_coeff_QL_Ir_non_sg}). 

Let us first study the $\gamma_\rmP \neq 2$ case. Evidently, $D(L,I_r)$ scales as ${\left|\Psi_{\rmP\ell_r}\right|}^2$, i.e., as $a^{2\left(2-\gamma_\rmP\right)}$ for $\gamma_\rmP < 3$ and as $a^{-2}$ for $\gamma_\rmP > 3$. If the density $\rho_0(r)\sim r^{-\gamma}$ with $0<\gamma<2$, then $\Omega_r = \partial E/\partial I_r \sim a^{-\gamma/2}$ and $f_0\sim a^{\gamma/2-3}$ (see section~\ref{sec:eqbm}). On the other hand, if $\rho_0(r)\sim r^{-\beta}$ with $\beta>3$, then $\Omega_r\sim a^{-3/2}$ and $f_0\sim a^{3/2-\beta}$. This implies that for $\rho_0(r)\sim r^{-\gamma}$ with $0<\gamma<2$, $\partial f_0/\partial I_r\sim \Omega_r \partial f_0/\partial E \sim a^{\gamma-5}$ and for $\rho_0(r)\sim r^{-\beta}$ with $\beta > 3$, $\partial f_0/\partial I_r\sim a^{1-\beta}$.

Let us now plug in the above scalings in the steady state condition given by equation~(\ref{steady_state_cond_const_flux}). If a system with $\rho_0(r)\sim r^{-\gamma}$ ($0<\gamma<2$) resides within a perturbing mass with $\rho_\rmP(r)\sim r^{-\gamma_\rmP}$ and $\gamma_\rmP<3$ that is fluctuating with a temporal power spectrum (Fourier transform of the temporal correlation), $\calC_{\omega}\left(\ell_r\Omega_r\right)\sim {\left(\ell_r\Omega_r t_\rmc\right)}^{-n_\gamma}$, then it tends to relax towards a steady state characterized by equation~(\ref{steady_state_cond_const_flux}), which implies the following relation between $\gamma$ and $\gamma_\rmP$:

\begin{align}
&\textstyle{a}^{\textstyle{\dfrac{n_\gamma\gamma}{2} + \gamma - 5 + 2\left(2-\gamma_\rmP\right)}} = {\rm constant} \nonumber\\
&\implies \gamma = \dfrac{1 + 2\gamma_\rmP}{1+\dfrac{n_\gamma}{2}}.
\label{steady_state_cond_const_flux_gamma}
\end{align}
If, on the other hand, the system is characterized by $\rho_0(r)\sim r^{-\beta}$ with $\beta>3$, and the perturbing mass with $\rho_\rmP(r)\sim r^{-\gamma_\rmP}$ and $0<\gamma_\rmP<2$ is fluctuating with a temporal power spectrum $\calC_{\omega}\left(\ell_r\Omega_r\right)\sim {\left(\ell_r\Omega_r t_\rmc\right)}^{-n_\beta}$, then the steady state condition of equation~(\ref{steady_state_cond_const_flux}) predicts the following relation between $\beta$ and $\gamma_\rmP$:

\begin{align}
&\textstyle{a}^{\textstyle{\dfrac{3n_\beta}{2} + 1 -\beta + 2\left(2-\gamma_\rmP\right)}} = {\rm constant} \nonumber\\
&\implies \beta = 5 - 2\gamma_\rmP + \frac{3n_\beta}{2}.
\label{steady_state_cond_const_flux_beta}
\end{align}

For $\gamma_\rmP = 2$, the diffusion coefficient scales logarithmically with $a$ and can therefore be modeled as a constant $D_0$ for all practical purposes. The QLDE is then a one-dimensional diffusion equation in $I_r$ with a constant diffusion coefficient, the self-similar solution to which is simply $f_0\left(I_r,\tau\right)\sim \exp{\left[-I^2_r/2\sigma^2_{I_r}\left(\tau\right)\right]}/\sqrt{2\pi \sigma^2_{I_r}\left(\tau\right)}$ with $\sigma^2_{I_r}\left(\tau\right) = \sigma^2_{I_r}\left(\tau=0\right) + 2 D_0\tau$.

\subsubsection{Inner halo}

Now we discuss how different parts of the halo develop different density log-slopes through quasilinear relaxation. Let the initial profile of the halo be $r^{-\gamma_\rmP}$ with $\gamma_\rmP \geq 0$, and let it be fluctuating with a generic temporal correlation $\calC_t$ such that $\calC_{\omega}\left(\ell_r\Omega_r\right)\sim {\left(\ell_r\Omega_r t_\rmc\right)}^{-n_\gamma}$ with $n_\gamma \geq 0$. Let the halo gravitationally accrete matter from outside with an arbitrary distribution. This newly accreted matter would now be heated by the fluctuating halo, which acts as the perturber. If the accreted matter develops a density profile $\rho_0(r)\sim r^{-\gamma}$ with $0<\gamma<2$ in the steady state, then we have from the above equation~(\ref{steady_state_cond_const_flux_gamma}) that

\begin{align}
\gamma = \dfrac{1 + 2\gamma_\rmP}{1 + \dfrac{n_\gamma}{2}}.
\label{steady_state_cond_const_flux_gamma_gammap0}
\end{align}
While the accreted matter is growing the $r^{-\gamma}$ cusp, the halo would be accreting more matter. If the rate of relaxation is higher than the accretion rate, then the halo would keep growing the $r^{-\gamma}$ cusp. Once the halo has grown to a critical mass, however, the accretion rate would exceed the rate of relaxation and the density log-slope would change. This sets the scale radius of the halo, inside (beyond) which virialization occurs faster (slower) than accretion.

\subsubsection{Outer halo}

If the $\rho_0(r) \sim r^{-\gamma}$ inner halo now accretes more matter, then this newly accreted material is perturbed and heated by the fluctuating inner halo. If the accreted matter develops the outer halo with a $\rho_0(r)\sim r^{-\beta}$ profile ($\beta > 3$) this way, then we have $\gamma_\rmP = \gamma$ in the expression for steady state $\beta$ in equation~(\ref{steady_state_cond_const_flux_beta}). The inner halo acts as the perturber in the assembly of the outer halo. Upon substituting the expression for steady state $\gamma$ from equation~(\ref{steady_state_cond_const_flux_gamma_gammap0}) in equation~(\ref{steady_state_cond_const_flux_beta}) (note that $\gamma_\rmP = \gamma$ here), we obtain the following steady state value of $\beta$:

\begin{align}
\beta &= 5 - 2\gamma + \frac{3n_\beta}{2} \nonumber\\
& = \dfrac{3}{1 + \dfrac{n_\gamma}{2}} \left[1 - \dfrac{4\gamma_\rmP}{3} + \dfrac{5n_\gamma}{6}\right] + \dfrac{3 n_\beta}{2}.
\label{beta_s}
\end{align}

Since the enclosed mass of the halo must be finite at $r\to \infty$, the density must fall off as $r^{-\beta}$ at large $r$ with $\beta \gtrsim 3$. This condition, together with the above relation, constrains the value of $\gamma$ to $\gamma \lesssim 1 + 3 n_\beta/4$. And this, together with equation~(\ref{steady_state_cond_const_flux_gamma_gammap0}), constrains the value of $\gamma_\rmP$ to $\gamma_\rmP \lesssim \left[n_\gamma + \left(3 n_\beta/2\right)\left(1+n_\gamma/2\right) \right]/4$. Equation~(\ref{beta_s}) puts a stringent constraint on the possible combinations of $\gamma$ and $\beta$ for a halo. If the inner halo fluctuates faster than the outer halo can assemble and relax, then the fluctuations can be modeled as a white noise and $n_\beta$ is small, implying that $\beta \approx 5 - 2\gamma$ in the steady state. Both the NFW profile with $\left(\gamma,\beta\right) = (1,3)$ and the Plummer sphere with $\left(\gamma,\beta\right) = \left(0,5\right)$ satisfy this condition, but the Hernquist sphere with $\left(\gamma,\beta\right) = (1,4)$ does not.

\subsection{Time-dependent solution}\label{sec:num_sol}

How is the above double power-law profile established? To answer this question, we solve the QLDE given by equation~[\ref{QLDE_Ir}] as an initial value problem, simultaneously with the Poisson equation.

\subsubsection{Inner halo}

If the density $\rho_0(r)$ of matter accreted in the pre-assembled halo scales as $r^{-\gamma}$ with $0<\gamma<2$, then the diffusion coefficient scales as $a^{2\left(2-\gamma_\rmP\right)}$, with $a\sim I_r^{2/\left(4-\gamma\right)}$, and the QLDE can be rewritten as

\begin{align}
\frac{\partial f_0}{\partial \tau} = \frac{\partial}{\partial \calI_r}\left(\calI_r^{\textstyle{\dfrac{2-\gamma_\rmP + \frac{n_\gamma\gamma}{4}}{1-\frac{\gamma}{4}}}}\frac{\partial f_0}{\partial \calI_r}\right),
\label{QLDE_Ir_gamma}
\end{align}
where we have defined $\tau = D_0\left(L\right) t/I^2_0 = t/t_{\rm diff}$, $\calI_r = I_r/I_0$, and $D\left(L,I_r\right) = D_0\left(L\right)\calI_r^{\left(2-\gamma_\rmP + n_\gamma\gamma/4\right)/\left(1 - \gamma/4\right)}$, with $I_0$ a characteristic radial action ($\sim \sqrt{G M_0 r_\rms}$) and $t_{\rm diff} = I^2_0/D_0\left(L\right)$ the diffusion timescale. For an initial value problem with constant $\gamma$, the above equation can be solved using the method of Green's function and a self-similar solution for the Green's function, as detailed in Appendix~C.2 of \citep[][]{Banik.etal.24}. However, in course of the quasilinear evolution, $\gamma$ does not remain constant. Once the pre-assembled halo accretes enough matter, $\gamma$ substantially changes. The QLDE evolves $f_0$, which alters the radial dependence of $\rho_0$ and $\Phi_0$ and therefore the value of $\gamma$. Hence, the QLDE must be rewritten with a time-evolving $\gamma$ as follows:

\begin{align}
\frac{\partial f_0}{\partial \tau} = \frac{\partial}{\partial \calI_r}\left(\calI_r^{\textstyle{\dfrac{2-\gamma_\rmP + \frac{n_\gamma\gamma\left(\tau\right)}{4}}{1-\frac{\gamma\left(\tau\right)}{4}}}}\frac{\partial f_0}{\partial \calI_r}\right).
\label{QLDE_Ir_gamma_t}
\end{align}

Let us now study how $\gamma$ evolves. We assume that $f_0\sim \calI_r^{-\kappa_0}$ initially, with arbitrary $\kappa_0$. The QLDE causes the gradual diffusion of $f_0$. Meanwhile, the Poisson equation~(\ref{Poisson_eq_rad}) dictates the evolution of $(\rho_0,\Phi_0)$ and therefore $\gamma$, which modifies the diffusion coefficient. If the density scales as $r^{-\gamma(\tau)}$ with $0 < \gamma(\tau) < 2$, the potential scales as $r^{2-\gamma(\tau)}$ according to the Poisson equation. Then $\calI_r$ scales as ${\left(\Psi_\rmc - \calE\right)}^{\,\left[2-\gamma(\tau)/2\right]\big/\left[2-\gamma(\tau)\right]}$ with $\Psi_\rmc = - \Phi_\rmc$. The density is given by 

\begin{align}
&\rho_0\left(\Psi_0\right) \sim \int_0^{\Psi_0} \rmd \calE\, \sqrt{2\left(\Psi_0-\calE\right)}\, f_0\left(\calI_r\left(\calE,\gamma\right)\right),
\label{rho_psi}
\end{align}
with $\Psi_0 = -\Phi_0$. The Poisson equation implies $\rho_0 \sim {\left(\Psi_\rmc - \Psi_0\right)}^{-\gamma/\left(2-\gamma\right)}$ and therefore reduces the above equation to

\begin{align}
{\Psi'_0}^{-\dfrac{\gamma\left(\tau\right)}{2 - \gamma\left(\tau\right)}} \sim \int_{\Psi'_0}^{\Psi_\rmc} \rmd \calE'\, \sqrt{2\left(\calE' - \Psi'_0\right)}\, f_0\left(\calE',\gamma\left(\tau\right)\right),
\label{vir_eff}
\end{align}
with $\Psi'_0 = \Psi_\rmc - \Psi_0$ and $\calE' = \Psi_\rmc - \calE$. If we now assume that $f_0\sim \calI_r^{-\kappa\left(\tau\right)}$ due to diffusion (this is a reasonable assumption over a large range in $\calI_r$ provided $f_0$ is initialized as $\calI_r^{-\kappa_0}$), then, using the fact that $\calI_r$ scales as ${\left(\Psi_\rmc - \calE\right)}^{\,\left[2-\gamma(\tau)/2\right]\big/\left[2-\gamma(\tau)\right]}$ in the above equation, we obtain that $\gamma(\tau) = {\left[2\left(2\kappa(\tau)-3\right)\right]}/{\left[\kappa(\tau)-1\right]}$. Noting that $\partial\gamma/\partial\tau = \left[2/{\left(\kappa-1\right)}^2\right]\partial\kappa/\partial\tau$ and $\partial\kappa/\partial\tau = -\partial/\partial\tau\left(\partial \ln f_0/\partial\ln\calI_r\right) = -\partial/\partial\ln\calI_r\left(\partial \ln f_0/\partial \tau\right)$, and using equation~(\ref{QLDE_Ir_gamma_t}), we find that

\begin{align}
\frac{\partial\gamma}{\partial\tau} &= 2\,{\left(1+\frac{n_\gamma}{2}\right)}^2\, \frac{6-\gamma}{4-\gamma}\left(\gamma - \frac{1+2\gamma_\rmP}{1+n_\gamma/2}\right)\left(\gamma - \frac{2\gamma_\rmP}{1+n_\gamma/2}\right)\nonumber\\
&\times \calI_r^{\dfrac{2\left[\gamma\left(1+\frac{n_\gamma}{2}\right)-2\gamma_\rmP\right]}{4-\gamma}}.
\label{dgamma_dt}
\end{align}
This is an effective model for virialization that tells us how the density-potential pair readjust to the gradually diffusing $f_0$ and the log-slope evolves with time (at each $\calI_r$). The virialization timescale is a function of $\calI_r$. As $\gamma$ decreases (increases), the diffusion coefficient in equation~(\ref{QLDE_Ir_gamma_t}) becomes a shallower (steeper) function of $\calI_r$, which drives $f_0$ towards a shallower (steeper) function with smaller (larger) $\kappa$ and $\gamma$. This process of relaxation continues until $\gamma$ reaches a steady state. It is evident that $\gamma = \gamma_\rmu = \left(1+2\gamma_\rmP\right)/(1+n_\gamma/2)$ and $\gamma = \gamma_\rms = 2\gamma_\rmP/(1+n_\gamma/2)$ are both fixed points of the above equation. Of these, $\gamma_\rmu$ is the true constant flux steady state solution to the QLDE, for which the RHS of the QLDE (equation~[\ref{QLDE_Ir_gamma_t}]) is zero for all $\calI_r$ (see section~\ref{sec:steady_state}). It is, however, an unstable fixed point of $\gamma$ evolution (equation~[\ref{dgamma_dt}]), as demonstrated by a stability analysis. The stable fixed point is rather $\gamma_\rms$. This is because $\partial \gamma/\partial \tau < 0$ for $\gamma_\rms < \gamma < \gamma_\rmu$ and positive otherwise. Therefore, the only way to achieve a steady state (at least approximately) in both $f_0$ and $\gamma$ is to have $\gamma_\rms \approx \gamma_\rmu$, which is attained for $n_\gamma \gtrsim 2$, i.e., red noise, and for $\gamma_0 = \gamma(0) \lesssim \gamma_\rmu$. Then, $\gamma_\rms\approx 4\gamma_\rmP/n_\gamma$ is a stable attractor. If $n_\gamma \approx 4\gamma_\rmP$, then $\gamma_\rmP \gtrsim 0.5$, and the attractor is approximately $1$. For finite values of $\gamma_\rmP$ and $n_\gamma$, though, $f_0$ only reaches a quasi-steady state, since $\gamma_\rmu$ is close to but not exactly equal to $\gamma_\rms$.

\begin{figure*}
\centering
\includegraphics[width=1\textwidth]{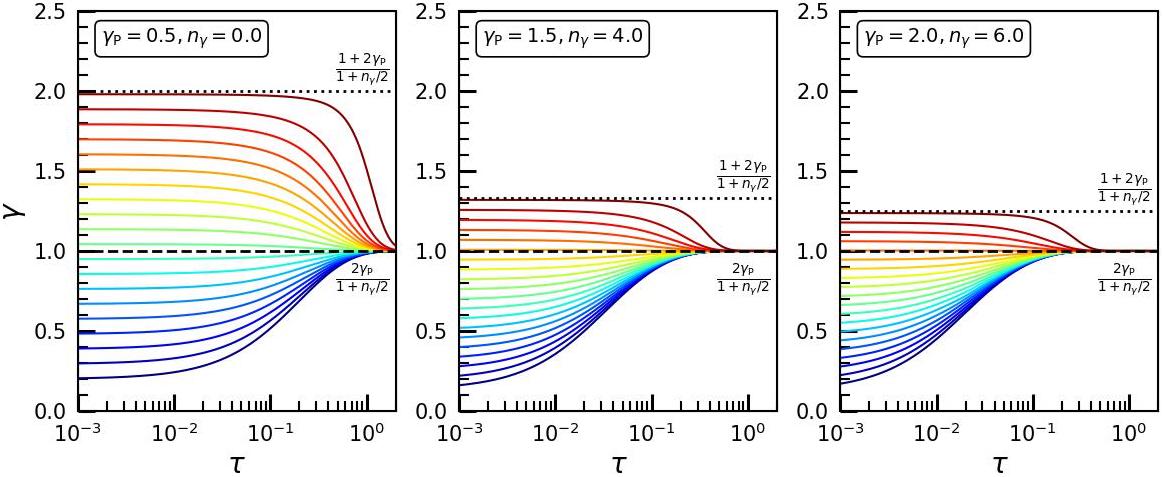}
\caption{Evolution of the inner log-slope $\gamma$ of a relaxing halo as a function of time $\tau=t/t_{\rm diff}$ ($t_{\rm diff} = I^2_0/D_0$) for different values of $\gamma_0=\gamma(\tau=0)$, $\gamma_\rmP$ and $n_\gamma$ as indicated, obtained by solving equation~(\ref{dgamma_dt}) at $\calI_r = I_r/I_0 = 1$. Dashed (dotted) line indicates the stable (unstable) fixed point. We adopt $\gamma_\rmP$ and $n_\gamma$ such that the stable fixed point is equal to $1$.}
\label{fig:gamma_vs_t_diff_gamma0}
\end{figure*}

We evolve $\gamma$ at $\calI_r = I_r/I_0 = 1$ using equation~(\ref{dgamma_dt}), for different initial conditions and for combinations of $\gamma_\rmP$ and $n_\gamma$ such that the stable fixed point $\gamma_\rms = 2\gamma_\rmP/(1+n_\gamma/2)$ is equal to $1$. We plot the resulting $\gamma$ as a function of $\tau$ in Fig.~\ref{fig:gamma_vs_t_diff_gamma0}. As expected, the trajectories diverge from the unstable fixed point, $\gamma_\rmu$, and converge towards the stable one, $\gamma_\rms$. Although, for $n_\gamma = 0$ and $\gamma_\rmP = 0.5$, $\gamma$ reaches the steady state $\gamma_\rms$, it is significantly different from $\gamma = \gamma_\rmu$, the constant flux steady state for $f_0$. As $n_\gamma$ increases, though, the stable and unstable points approach each other and both $f_0$ and $\gamma$ attain a quasi-steady state. In this limit, the range of $\gamma_\rms < \gamma_0 < \gamma_\rmu$ for which $\gamma$ approaches $\gamma_\rms$ becomes negligible, and only for $\gamma_0 < \gamma_\rms$ does $\gamma$ effectively approach $\gamma_\rms$. Hence, while $\gamma_\rms$ is technically a stable attractor, it effectively behaves as a neutral one. Starting below $\gamma_\rms$, $\gamma$ increases and settles at $\gamma_\rms$, but does not go all the way up to $\gamma_\rmu$, the true steady state for $f_0$, which reflects the self-limiting nature of collisionless relaxation.

What is the timescale of relaxation of the inner halo? The mean dynamical time at $r_\rms$ is $t_{\rm dyn}\sim \sqrt{r^3_\rms/G M}$. The relaxation time is the timescale of quasilinear diffusion, $t_{\rm diff} = I^2_0/D_0$. Since $D_0\sim {\left(G M /r_\rms\right)}^2 t_\rmc$ (recall from equation~[\ref{diff_coeff_QL_Ir_non_sg}] that the diffusion coefficient has the units of potential squared times time), $t_\rmc$ being the correlation time of fluctuations, and $I_0 \sim \sqrt{G M r_\rms}$, $t_{\rm diff}$ turns out to be $t^2_{\rm dyn}/t_\rmc$. The ratio of the quasilinear diffusion timescale to the dynamical time is therefore equal to

\begin{align}
\frac{t_{\rm diff}}{t_{\rm dyn}} \sim \frac{t_{\rm dyn}}{t_{\rmc}}.
\label{timescale_sep}
\end{align}
Since relaxation to the quasi-steady state requires red noise as shown above, the correlation time $t_\rmc$ is of order $t_{\rm dyn}$, implying that $t_{\rm diff}$ is of order $t_{\rm dyn}$. Hence, the inner halo relaxes or virializes over a dynamical time. This suggests that there is not much of a timescale separation in the inner halo as one would ideally like to have in a quasilinear framework. A proper analysis of collisionless relaxation would thus require one to go beyond QLT and construct an effective theory for violent relaxation, something that is beyond the scope of this paper but we plan to address in future work.

%\begin{figure}
%\centering
%\includegraphics[width=1\textwidth]{gamma_vs_t.jpeg}
%\caption{Same as Fig.~\ref{fig:gamma_vs_t_long_time} but zoomed into earlier times, $\tau<0.6$. Note that the system spends substantial time near the quasi-steady state of $\gamma\approx 1$.}
%\label{fig:gamma_vs_t}
%\end{figure}

\subsubsection{Outer halo}

\begin{figure*}
\centering
\includegraphics[width=1\textwidth]{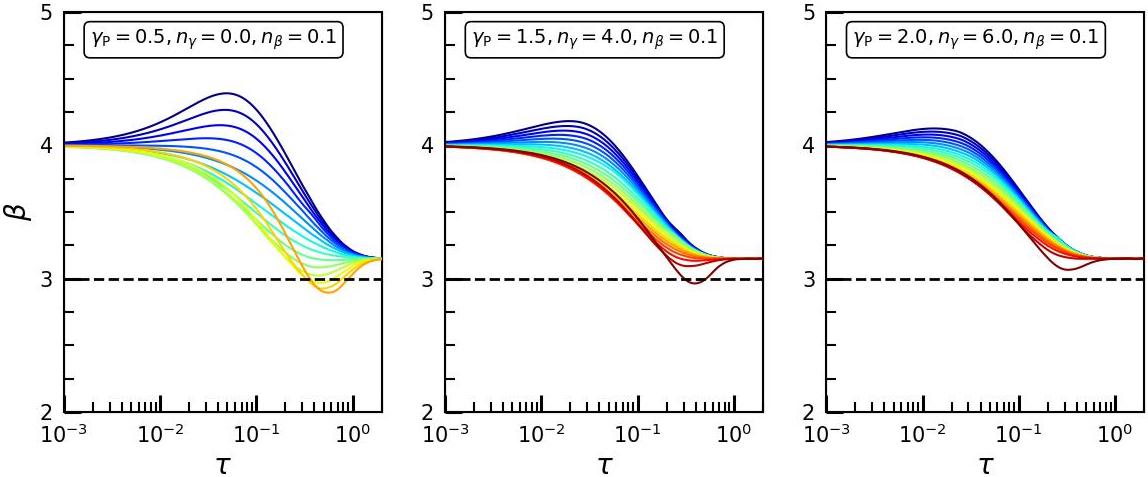}
\caption{Evolution of the outer log-slope $\beta$ as a function of $\tau=t/t_{\rm diff}$ ($t_{\rm diff} = I^2_0/D_0$), obtained by solving equations~(\ref{dgamma_dt}) and (\ref{dbeta_dt}), for the combinations of $\gamma_0$, $\gamma_\rmP$ and $n_\gamma$ adopted in Fig.~\ref{fig:gamma_vs_t_diff_gamma0} such that $\gamma_\rms = 1$. We adopt $n_\beta = 0.1$, and restrict ourselves to $\gamma_0 < 1.5\left(1+n_\beta/2\right)$ so that the fixed point $\beta_\rms = 5 - 2\gamma_\rms + 3n_\beta/2$ is stable. As $\gamma$ approaches the stable fixed point $\gamma_\rms = 1$ in Fig.~\ref{fig:gamma_vs_t_diff_gamma0}, $\beta$ approaches $\beta_\rms = 3.15$.}
\label{fig:beta_vs_t}
\end{figure*}

\paragraph{Quasilinear diffusion due to inner halo\textemdash}\label{sec:outer_halo_QLDE}

As the halo builds up, it accretes more matter that is perturbed by the fluctuating inner halo. This forms the outer halo whose density falls off as $r^{-\beta}$ with $\beta>3$. The quasilinear diffusion coefficient then scales as $a^{2\left(2-\gamma\right) + 3 n_\beta/2}\sim I_r^{4\left(2-\gamma\right) + 3 n_\beta}$, since $a\sim I^2_r$, and $\gamma_\rmP$ is now equal to $\gamma$, the log-slope of the inner halo. Consequently, $f_0$ of the outer halo evolves via the following QLDE:

\begin{align}
\frac{\partial f_0}{\partial \tau} = \frac{\partial}{\partial \calI_r} \left( \calI^{\,4\left(2-\gamma(\tau)\right) + 3 n_\beta}_r \frac{\partial f_0}{\partial \calI_r} \right).
\label{QLDE_Ir_beta_t}
\end{align}
Starting from the initial scaling of $\calI_r^{-\eta_0}$, $f_0\left(\calI_r\right)$ gets progressively shallower due to quasilinear diffusion. Since $\beta > 3$, the potential scales as $-r^{-1}$ according to the Poisson equation, and $\calI_r$ as $\calE^{-1/2}$ ($\calE = \left|E\right|$). The density is given by equation~(\ref{rho_psi}) with $\calI_r \sim {\calE}^{-1/2}$. Since $\rho_0\sim r^{-\beta}$ and $\Psi_0\sim r^{-1}$, we have that $\rho_0\sim \Psi_0^{\,\beta}$, i.e.,

\begin{align}
\Psi_0^{\,\beta\left(\tau\right)} \sim \int_0^{\Psi_0} \rmd \calE\, \sqrt{2\left(\Psi_0-\calE\right)}\, f_0\left(\calE,\gamma\left(\tau\right)\right).
\label{beta_eq}
\end{align}
Therefore, the evolution of $\gamma$ dictates that of $\beta$. Assuming that $f_0(\calI_r)\sim \calI_r^{-\eta(\tau)}$ and using the fact that $\calI_r\sim \calE^{-1/2}$, we find that $\beta(\tau) = \left(\eta(\tau)+3\right)/2$. Noting that $\partial\beta/\partial\tau = (1/2)(\partial\eta/\partial\tau)$ and $\partial\eta/\partial\tau = -\partial/\partial\tau\left(\partial \ln f_0/\partial\ln\calI_r\right) = -\partial/\partial\ln\calI_r\left(\partial \ln f_0/\partial \tau\right)$, and using equation~(\ref{QLDE_Ir_beta_t}), we find that

\begin{align}
\frac{\partial\beta}{\partial\tau} &= -8\left(\frac{3}{2}\left(1+\frac{n_\beta}{2}\right)-\gamma\right)
\left(\beta-\frac{3}{2}\right)
\left(\beta - \left(5 - 2\gamma + \frac{3 n_\beta}{2}\right)\right)\nonumber\\
&\quad\times \calI_r^{\displaystyle 4\left[\frac{3}{2}\left(1+\frac{n_\beta}{2}\right)-\gamma\right]}.
\label{dbeta_dt}
\end{align}
There are two possible fixed points of the above equation: $\gamma = 1.5(1 + n_\beta/2)$ and $\beta = 5 - 2\gamma + 3n_\beta/2$ (recall that $\beta>3$ and therefore cannot be $3/2$). The constant flux steady state solution to the QLDE (equation~[\ref{QLDE_Ir_beta_t}]) is $\beta_\rms$. Therefore, if $\gamma = 1.5(1 + n_\beta/2)$, then we also require $\beta = 5 - 2\gamma + 3n_\beta/2 = 2$, to ensure a steady state for both $f_0$ and $\beta$. However, $\beta$ is the outer log-slope and must exceed $3$ for finite total mass, which implies that $\gamma = 1.5(1 + n_\beta/2)$ is not a valid steady state solution. The only valid steady state is therefore $\beta_\rms = 5 - 2\gamma_\rms + 3n_\beta/2$, where $\gamma_\rms = 2\gamma_\rmP/(1+n_\gamma/2)$ is the stable fixed point in $\gamma$. Since $\gamma_\rms < 1.5(1 + n_\beta/2)$ (so that $\beta_\rms > 3$), $\beta_\rms$ is a stable fixed point. This is because $\partial \beta/\partial \tau > 0$ $(<0)$ if $\beta < \beta_\rms$ $(\beta > \beta_\rms)$. As $\gamma$ decreases (increases), the diffusion coefficient in equation~(\ref{QLDE_Ir_gamma_t}) becomes a steeper (shallower) function of $\calI_r$, which drives $f_0$ towards a steeper (shallower) function with larger (smaller) $\eta$ and $\beta$. Since $\beta_\rms > 3$, $\gamma_\rms < 1 + 3 n_\beta/4$ is a strict requirement, which in turn necessitates $\gamma_\rmP < n_\gamma(1 + 3n_\beta/4)/4$.

We evolve $\beta$ at $\calI_r = 1$ using equation~(\ref{dbeta_dt}) and $\gamma(\tau)$ obtained from equation~(\ref{dgamma_dt}), assuming $\beta_0 = \beta(0) = 4$, $n_\beta = 0.1$, $\gamma_0 < 1.5(1 + n_\beta/2)$ (such that $\beta_\rms$ is stable), and the same combination of $\gamma_\rmP$ and $n_\gamma$ as in Fig.~\ref{fig:gamma_vs_t_diff_gamma0} (such that $\gamma_\rms=1$). We plot the resulting $\beta$ as a function of $\tau$ in Fig.~\ref{fig:beta_vs_t}. If $\gamma_0 < 0.5$, then $\left.\partial\beta/\partial\tau\right|_{\tau\to 0^+} > 0$ and $\beta$ initially increases, since $\gamma_0 < 1.5(1 + n_\beta/2)$ and $\beta_0 < 5 - 2\gamma_0 + 3 n_\beta/2$. For these values of $\gamma_0$, $\beta$ passes through a local maximum with $\beta = 5 - 2\gamma + 3 n_\beta/2$. Eventually, $\beta$ decreases and approaches the stable fixed point $\beta_\rms = 5 - 2\gamma_\rms + 3 n_\beta/2 = 3.15$ as $\gamma$ approaches $\gamma_\rms = 2\gamma_\rmP/\left(1+n_\gamma/2\right) = 1$. For $\gamma_0 > 0.5$, it does so after passing through a local minimum. All in all, $\beta$ approaches $\beta_\rms$ as long as $\gamma_0 < 1.5(1 + n_\beta/2)$.

\paragraph{Accretion and spherical collapse \textemdash}\label{sec:accretion}

We have seen how the outer halo develops an $r^{-\beta}$ profile, with $\beta = 5 - 2\gamma + 3n_\beta/2$, under the fluctuating forces exerted by the inner halo on the accreted matter. Let us now try to derive the outer fall-off without the quasilinear approximation, using the spherical collapse theory for the radial motion of accreted shells.

Spherically symmetric radial infall of a bound shell of matter with radius $r(t)$ onto the halo yields the following oscillating solution: $r = r_\rma\left(1-\cos{\theta}\right)$ and $t = t_\rma\left(\theta - \sin{\theta}\right)$, with $r_\rma = GM/2\left|E\right|$ the apocentric radius, $E$ the energy of the shell, $t_\rma$ the orbital period and $r^3_\rma = G M t^2_\rma$. Here we have assumed that the mass enclosed within the shell varies slower than the shell oscillates, which is true sufficiently far from its apocenter. If $r_\rma \gg r_\rms$, then the shell spends most of its time beyond $r_\rms$, wherein $r(t)$ of an outgoing shell scales as $r(t) \approx r_\rms {\left(M/M_\rms\right)}^{1/3} {\left(t/t_\rms\right)}^{2/3}$, with $M_\rms = M(r_\rms)$ the mass enclosed within $r_\rms$ and $t_\rms$ the dynamical time at $r_\rms$. If the halo density profile scales as $\rho_0(r)\sim r^{-\beta}$ with $\beta > 3$ beyond $r_\rms$, then the enclosed mass profile is given by $M(r) = M_\rms + \left(M_\infty - M_\rms\right)\left[1 - {\left(r/r_\rms\right)}^{3-\beta}\right]$ with $M_\infty = M(r\to \infty)$. This implies that the mass within a shell with radius $r(t)\approx r_\rms {\left(M/M_\rms\right)}^{1/3} {\left(t/t_\rms\right)}^{2/3}$ evolves as

\begin{align}
m = m_\infty - \left(m_\infty - 1\right) m^{{1 - {\beta}/{3}}}\, \tau^{\,{2\left(1 - {\beta}/{3}\right)}},
\label{mass_evol}
\end{align}
where $m = M(t)/M_\rms$, $m_\infty = M_\infty/M_\rms$ and $\tau = t/t_\rms$. Hence, in a halo with $\beta$ nearly equal to (but slightly larger than) $3$, the mass enclosed within the shell barely changes as it moves. For steeper profiles, the enclosed mass rapidly rises and saturates to $m_\infty$ as the shell moves beyond $r_\rms$. This rapid increase in the enclosed mass binds the shell more tightly, reducing its (negative) energy and apocenter. The accreted shells keep getting more and more bound and thus pile up in the outer halo until $\beta$ reaches $\sim 3$, in which case the mass enclosed within a shell (and also its energy) remains nearly constant as it moves in the outer halo (according to equation~[\ref{mass_evol}]). This proves that the outer density profile of a quasi-steady halo must scale as $\sim r^{-3}$. 

We have seen that the accretion and quasilinear relaxation of matter in an $r^{-\gamma}$ inner halo establishes an $r^{-\beta}$ outer halo with $\beta = 5 - 2\gamma + 3n_\beta/2$ in the quasi-steady state. Spherical collapse theory for matter orbiting in the outer halo suggests that a quasi-steady halo requires $\beta \approx 3$. These two results together imply that, in the steady state, $\beta = 5 - 2\gamma + 3n_\beta/2 \approx 3$, i.e., $\gamma \approx 1 + 3 n_\beta/4$. If the inner halo fluctuates (virializes) on timescales much shorter than that of the accretion and relaxation of the outer halo, as is typically the case, then the forces exerted by the fluctuating inner halo can be approximated as a white noise, i.e., $n_\beta$ can be assumed to be small. This dictates that $\gamma \approx 1$ is a quasi-steady solution. In other words, an NFW-like profile with $\gamma\approx 1$ and $\beta \approx 3$ is a quasi-steady state of collisionless relaxation. And, since the outer halo relaxes to an $r^{-3}$ profile under white noise fluctuations (small $n_\beta$), equation~(\ref{timescale_sep}) dictates that $t_{\rm diff}/t_{\rm dyn} \sim t_{\rm dyn}/t_\rmc \gg 1$, implying that timescale separation is a valid assumption and QLT is an appropriate tool for describing the relaxation of the outer halo.

What is the physics behind the establishment of an NFW halo? Ongoing accretion (or any other gravitational perturbation) imparts energy to the inner halo. Subsequent virialization converts the excess kinetic energy into potential energy, which is a consequence of the negative specific heat of self-gravitating systems (virial theorem). This cools the inner halo until it settles to an $r^{-1}$ profile, the coldest $r^{-\gamma}$ profile with $0\leq \gamma \leq 2$. The velocity dispersion of an isotropic inner halo scales as $r^{\gamma/2}$ for $0\leq \gamma \leq 1$ and as $r^{1-\gamma/2}$ for $1\leq \gamma \leq 2$ \citep[][]{Dehnen.93}, and is therefore the lowest at $r\to 0$ for $\gamma = 1$. As such, the formation of the $r^{-1}$ inner NFW cusp may be viewed as a consequence of the virial theorem. The inner region of a CDM halo tends to cool down. As the $r^{-1}$ inner cusp is established, accretion onto it with simultaneous relaxation forms an $r^{-3}$ outer halo in the quasi-steady state, since the mass enclosed within a spherical shell barely changes with time for such a radial fall-off. The establishment of the $r^{-1}$ inner cusp, however, requires a pre-assembled halo with $\gamma_\rmP \approx n_\gamma/4 \gtrsim 0.5$ (in the quasilinear framework). Whether the temporal correlation of fluctuations in the prompt cusp $(\gamma_\rmP = 1.5)$, around which \citet[][]{Delos.White.23} find the NFW halo to gradually assemble, satisfies the condition $n_\gamma\approx 4\gamma_\rmP = 6$, is a matter worthy of future investigation.

\subsection{Zero flux solution}\label{sec:zero_flux}

Rather than relaxing to a constant flux steady state discussed so far, part of the halo may relax to a zero flux steady state, wherein diffusion halts due to the erasure of energy gradients in the system. This amounts to the following trivial steady state condition:

\begin{align}
{\rm Flux} = -D\left(L,I_r\right)\frac{\partial f_0}{\partial I_r} = 0 \implies \frac{\partial f_0}{\partial I_r} = 0,
\label{steady_state_cond_zero_flux}
\end{align}
i.e., the DF is independent of $I_r$ or $E$. The corresponding density can still be a non-trivial function of $r$ due to the radial dependence of the escape velocity $\sqrt{2\left|\Phi_0\right|}$. The density $\rho_0$ can be obtained in terms of the galaxy potential $\Phi_0$ as follows:

\begin{align}
\rho_0 = 4\pi\int_0^{\Psi_0} \rmd \calE \sqrt{2\left(\Psi_0-\calE\right)}\, f_0 \sim \Psi^{3/2}_0,
\end{align}
with $\calE = -E$ and $\Psi_0 = -\Phi_0$. This reduces the Poisson equation~\ref{Poisson_eq_rad} to the following Lane-Emden equation of order $n=3/2$:

\begin{align}
\frac{1}{s^2}\frac{\rmd}{\rmd s}\left(s^2\frac{\rmd \psi}{\rmd s}\right) = -\psi^{3/2},
\label{Lane_Emden_eq_1.5}
\end{align}
with $\psi = \Psi_0/\Psi_\rms$ and $s = r/r_\rms$, $\Psi_\rms$ and $r_\rms$ being the absolute value of the characteristic potential and the scale radius of the halo respectively. The above equation has two solutions for two sets of boundary condition. If $\psi$ tends to a constant and $\rmd\psi/\rmd s\to 0$ at $s\to 0$, then both $\psi$ and $\rho_0$ follow a cored profile with compact support (i.e., truncated at some radius). The halo profile therefore harbors a central core with a smooth roll-over of the outer log-slope, but is truncated. On the other hand, if $\psi\sim s^{-1}$ at $s\to 0$, then $\psi$ scales as $s^{-1}$ for a large range in $s$ before falling off to zero at some radius. The corresponding $\rho_0$ scales as $s^{-3/2}$ before truncation. This might happen if the halo centers around a massive compact object with a density profile falling off more steeply than $r^{-3}$, or if the halo assembles around a black hole. In fact, this $r^{-3/2}$ profile emerges naturally as a self-similar solution of the infall of collisionless fluid onto a black hole in the spherical collapse model of \citep[][]{Bertschinger.85}, as long as it does not undergo shell-crossing. Note that the $r^{-3/2}$ cusp grows around a compact perturber that is impulsively introduced, which is very different from the formation of a much steeper density cusp around an adiabatically growing black hole \citep[][]{Gondolo.Silk.99}. Although the $r^{-3/2}$ scaling of $\rho_0$ is the same as in the prompt cusp that appears in the early stage of halo formation \citep[][]{Delos.White.23}, the prompt cusp is quantitatively different from this. Here, the $r^{-3/2}$ cusp requires the presence of a central dense object, which is why the potential scales as $-r^{-1}$ around it. The potential of the prompt cusp, on the other hand, scales as $r^{1/2}$. 

\begin{figure}
\centering
\includegraphics[width=1\textwidth]{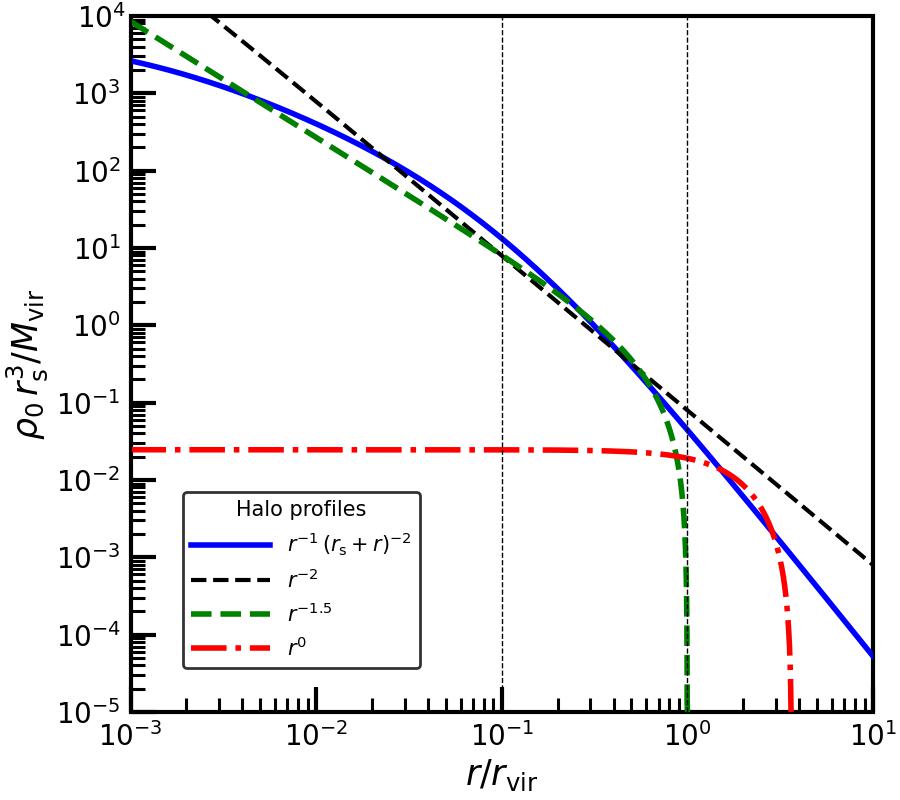}
\caption{Halo density $\rho_0$ (in units of $M_{\rm vir}/r^3_{\rms}$) as a function of radius $r$ (in units of $r_{\rm vir}$). The solid blue line indicates the NFW profile, the constant flux quasi-steady state. The dot-dashed red and dashed green lines respectively indicate the central core and $r^{-1.5}$ profiles, which are zero flux steady states obtained by numerically integrating the Lane-Emden equation~(\ref{Lane_Emden_eq_1.5}). The dashed black line indicates the isothermal sphere. The vertical dashed lines indicate the virial radius $r_{\rm vir}$ and the scale radius, $r_\rms$, assumed to be $0.1 r_{\rm vir}$. The profiles are normalized such that the virial mass of the NFW and isothermal sphere profiles is the same as the total mass of the other two.}
\label{fig:rho_halo_vs_r}
\end{figure}

Fig.~\ref{fig:rho_halo_vs_r} plots the cored and $r^{-3/2}$ profiles, obtained by numerically integrating equation~(\ref{Lane_Emden_eq_1.5}), as dot-dashed red and dashed green lines, the NFW density profile (equation~[\ref{rho_nfw}]) as a solid blue line, and the isothermal sphere as a dotted black line, as a function of $r/r_{\rm vir}$, where $r_{\rm vir}$ is the virial radius of the NFW halo (defined as the radius within which the mean halo density is $\sim 200$ times the critical density of the universe). The virial mass $M_{\rm vir} = M_0\left(r_{\rm vir}\right)$ of the NFW halo is equal to $4\pi \rho_\rmc r^3_\rms g(c)$ with $c = r_{\rm vir}/r_\rms$ the concentration parameter of the halo and $g(c)=\ln{\left(1+c\right)}-1/(1+c)$. We assume $r_\rms = 0.1 r_{\rm vir}$, i.e., $c=10$. The profiles have been normalized such that the mass enclosed within the $r^{-3/2}$ cusp matches that within the core as well as the virial masses of the NFW halo and the isothermal sphere. The zero flux solution is valid in the very central part of the halo. If it harbors (does not harbor) a central compact object, it develops an $r^{-1.5}$ cusp (a central core). Surrounding this, an NFW-like profile develops, as discussed in section~\ref{sec:quasilinear}.

\section{Discussion and summary}\label{sec:discussion_summary}

We have developed a self-consistent quasilinear theory for the collisionless relaxation of self-gravitating systems. Using this theory, we have shown that while the evolution of the fine-grained DF is described by the Vlasov equation, that of the coarse-grained DF $f_0$ is governed, under the quasilinear approximation, by a diffusion equation that we call the quasilinear diffusion equation (QLDE). It describes how the non-linear coupling of the linear fluctuations sourced by stochastic gravitational perturbations drives the secular evolution of $f_0$.

In this paper, we investigate the assembly of a halo via gravitational accretion and collisionless relaxation of the accreted matter. We use QLT to describe the evolution of $f_0$ of the accreted material (system) under stochastic perturbations of the pre-assembled halo (perturber). A key aspect of the theory is the dependence of the quasilinear diffusion coefficient not only on the perturbing potential but also on the mean potential of the system, which itself changes upon the evolution of its mean DF $f_0$. This self-consistency is a key aspect of our theory. In a way, this is an effective theory for virialization. We find that, when a pre-assembled halo with an $r^{-\gamma_\rmP}$ profile accretes matter on timescales comparable to the correlation time $t_\rmc$ of its fluctuations (modeled as a red noise with correlation $\calC_\omega \sim {(\omega t_\rmc)^{-n_\gamma}}$), the accreted material undergoes diffusive heating and settles into an $\sim r^{-\gamma}$ cusp with $\gamma \approx \gamma_\rms = 4\gamma_\rmP/n_\gamma$ as long as $n_\gamma \gtrsim 2$ and the initial value of $\gamma$ is smaller than $\gamma_\rms$ (the initial distribution is shallow or hot enough). This suggests that $\gamma \approx 4\gamma_\rmP/n_\gamma$ is something like a neutral equilibrium. As more matter gets accreted by the $r^{-\gamma}$ cusp, it relaxes to an $\sim r^{-\beta}$ outer profile with $\beta \approx 5 - 2\gamma$, under perturbations by the (fluctuating/virializing) $r^{-\gamma}$ cusp. On the other hand, spherical collapse theory predicts that a quasi-steady outer halo must settle to $\beta\approx 3$ in order that the mass enclosed within a radially moving shell is nearly time independent. QLT then dictates that $\gamma \approx (5-\beta)/2 \approx 1$, i.e., the $r^{-3}$ outer halo must have assembled via accretion and relaxation in an $r^{-1}$ inner halo. This suggests that the NFW profile is a self-consistent quasi-steady state of collisionless relaxation. The assembly of the $r^{-1}$ inner halo, however, requires, under the quasilinear approximation, that the pre-assembled halo harbors a steep enough profile with $\gamma_\rmP \approx n_\gamma/4 \gtrsim 0.5$. It would be interesting to see if the fluctuations of the prompt cusp \citep[][]{Delos.White.23}, for which $\gamma_\rmP = 1.5$, satisfy the condition $n_\gamma\approx 4\gamma_\rmP = 6$. If the halo harbors a (impulsively grown) central black hole, the innermost part develops an $r^{-1.5}$ cusp and if not, it forms an isothermal core.

The quasilinear analysis suggests that the NFW profile has the characteristics of a neutral equilibrium. Only matter accreted with profiles shallower than $r^{-1}$ can relax to an $r^{-1}$ cusp. This dependence on the initial conditions and the nature of the accretion flow shows that the NFW profile is probably not as universal as originally claimed. However, our current formalism suffers from an obvious caveat: it is based on the quasilinear approximation that assumes timescale separation, which is well-suited for the assembly of the outer halo but less so for that of the inner cusp. A more appropriate theory for halo formation would require going beyond the quasilinear regime, e.g., by (i) incorporating the modification of linear perturbations by non-linear mode coupling (weak turbulence theory), or (ii) developing an effective theory for violent relaxation. We plan to address some of this in future work.

Our approach towards modeling collisionless relaxation, while being fundamentally different from most previous attempts to explain the origin of the NFW profile, is similar to that of \citet[][]{Weinberg.01a,Weinberg.01b}, who solves the QLDE to study the relaxation of a halo perturbed by orbiting satellites. Contrary to our prediction, though, he obtains an Einasto-like and not the NFW profile in the quasi-steady state. We believe that the following factors are responsible for this discrepancy: (1) due to computational complexity, he does not study the evolution of initially cuspy profiles, and (2) he investigates the response of the halo to orbiting subhalos/satellites, a scenario different from the assembly of the halo that we concern ourselves with. In the scenario of \citet[][]{Weinberg.01a,Weinberg.01b}, the subhalos inspiral under dynamical friction, heat the host halo, and give rise to a cored, Einasto-like halo profile over time. It is possible that the NFW profile that initially forms via accretion and relaxation would transition to an Einasto-like rollover in the long run if we allowed for similar substructure perturbations. We leave a detailed investigation of this for future work. It would be interesting to see what halo profiles are predicted by the cosmological QLT of \citet[][]{Ma.Bertschinger.04}. Whether there exists a deep connection between the halo profiles and the small-scale matter power-spectrum \citep[][]{Ginat.etal.25,Nastac.etal.25,Nastac.etal.23} is a subject that deserves future investigation.

We have only looked for spherically symmetric and isotropic/ergodic solutions to the QLDE in this paper. There is, however, an entire landscape of distributions that satisfy the steady state condition obtained by putting the RHS of equation~(\ref{QL_diff_eq_gen}) to zero, with the diffusion tensor given by equation~(\ref{diff_coeff_QL}). This condition reduces the enormous landscape of steady state solutions allowed by the Vlasov equation to one with a much smaller measure. Instead of {\it any} positive definite function of the conserved quantities or actions as allowed by the Vlasov equation, now we have a {\it restricted} set of functions that follow the quasilinear equation. We have shown in this paper that the NFW profile emerges as a quasi-steady state of the collisionless relaxation of a spherical isotropic halo. Deviations from spherical symmetry and isotropy would of course give rise to very different equilibrium profiles. For example, it would be interesting to see if the exponential surface density profile that appears to be ubiquitous among disk galaxies \citep[][]{vdBosch.01,Elmegreen.Struck.13,Herpich.etal.17} emerges as an axisymmetric steady state of collisionless relaxation. Our formalism can also be used to study the modification of galaxy and halo profiles by baryonic feedback and secular evolution or dynamical friction \citep[][]{Freundlich.etal.20,Dekel.etal.21,Li.etal.23,Dattathri.etal.25}. And, last but not least, this work only serves as a stepping stone towards understanding the fascinating topic of violent relaxation. Much work is needed to understand the role of intrinsically non-linear effects such as particle trapping \citep[][]{Hamilton.24} in structure formation.

\begin{acknowledgments}
The authors are thankful to the Kavli Institute of Theoretical Physics (KITP), University of California, Santa Barbara, where much of the manuscript was prepared, and to the organizers and attendees of the workshop, ``Interconnections between the Physics of Plasmas and Self-gravitating Systems" at KITP, for insightful discussions. The authors are particularly grateful to Rimpei Chiba for providing valuable insights and to the anonymous referee for providing valuable suggestions that greatly improved the manuscript. The authors are also thankful to Martin Weinberg, Frank van den Bosch, Barry Ginat, Alex Schekochihin, Robert Ewart, Michael Nastac, Neal Dalal, Chris Hamilton, Scott Tremaine, Sten Delos, Nicholas Kokron and Matthew Kunz for stimulating discussions. This research is supported by the Bezos Member Fund and the Fund for Memberships in Natural Sciences at the Institute for Advanced Study, the National Science Foundation Award 2206607 at the Multi-Messenger Plasma Physics Center (MPPC), and Princeton University.
\end{acknowledgments}

\appendix

\section{Linear response theory}\label{App:lin_resp}

The linearized Vlasov equation given by the first of equations~(\ref{lin_CBE_Poisson}) can be solved in the angle-action $(\bw,\bI)$ space, in which case it reduces to

\begin{align}
\frac{\partial f_1}{\partial t} + {\bf \Omega} \cdot \frac{\partial f_1}{\partial \bw}  = \frac{\partial f_0}{\partial \bI} \cdot \frac{\partial H_1}{\partial \bw},
\label{lin_CBE_act_ang}
\end{align}
where ${\bf \Omega} = (\Omega_1, \Omega_2, \Omega_3)$ (in 3D) are the frequencies, given by
\begin{align}
{\bf \Omega} = \frac{\partial H_0}{\partial \bI}
\end{align}
It gets further simplified in the Fourier space of the angles. We expand $f_1$, $\Phi_1$ and $\Phi_\rmP$ as Fourier series in angles:

\begin{align}
f_1(\bw,\bI,t) &= \sum_{\boldell} \exp{\left[i \boldell \cdot \bw\right]}\, f_{1\boldell}(\bI,t),\nonumber \\
\Phi_1(\bw,\bI,t) &= \sum_{\boldell} \exp{\left[i \boldell \cdot \bw\right]}\, \Phi_{1\boldell}(\bI,t), \nonumber \\
\Phi_\rmP(\bw,\bI,t) &= \sum_{\boldell} \exp{\left[i \boldell \cdot \bw\right]}\, \Phi_{\rmP\boldell}(\bI,t).
\end{align}
This reduces equation~(\ref{lin_CBE_act_ang}) to the following evolution equation for $f_{1\boldell}$:

\begin{align}
\frac{\partial f_{1\boldell}}{\partial t} + i\boldell\cdot{\bf \Omega}f_{1\boldell} = i\boldell\cdot\frac{\partial f_0}{\partial \bI}\left(\Phi_{1\boldell} + \Phi_{\rmP\boldell}\right).
\label{lin_CBE_lmode}
\end{align}
Since we are interested in an initial value problem, we also take the Laplace transform in time:

\begin{align}
\Tilde{Q}(\bI,\omega) = \int_0^\infty \rmd t\, \exp{\left[i\omega t\right]}\, Q(\bI,t).
\end{align}
This reduces equation~(\ref{lin_CBE_lmode}) to the following equation for $\Tilde{f}_{1\boldell}(\bI,\omega)$:
 
\begin{align}
\Tilde{f}_{1\boldell}(\bI,\omega) = - \,\boldell \cdot \frac{\partial f_0}{\partial \bI} \,\frac{\Tilde{\Phi}_{1\boldell} + \Tilde{\Phi}_{\rmP\boldell}}{\omega - \boldell \cdot {\bf \Omega}} + \frac{i f_{1\boldell}(\bI,0)}{\omega - \boldell \cdot {\bf \Omega}},
\label{f1l_app}
\end{align}
with $f_{1\boldell}(\bI,0)$ the initial value of $f_{1\boldell}(\bI,t)$.

Now, we need to relate $\Phi_{1\boldell}$ to $f_{1\boldell}$ through the Poisson equation. The gravitational potential, $\Phi$, is related to the density, $\rho = \int \rmd^3v f$ by

\begin{align}
\Phi(\bx) = \int \rmd^3 x' \, U(\bx,\bx') \,\rho(\bx'),
\end{align}
with the pairwise interaction potential, $U(\bx,\bx') = -G/\left|\bx - \bx'\right|$. This implies that $\Tilde{\Phi}_{1\boldell}$ is related to $\Tilde{f}_{1\boldell}$ as follows:

\begin{align}
\Tilde{\Phi}_{1\boldell} (\bI) = {\left(2\pi\right)}^3 \sum_{\boldell'} \int \rmd \bI' \Psi_{\boldell \boldell'} (\bI,\bI') \Tilde{f}_{1\boldell'}(\bI'),
\label{Phi1l_f1l}
\end{align}
with

\begin{align}
&\Psi_{\boldell\boldell'} (\bI,\bI') \nonumber\\
&= \int \frac{\rmd^3 w}{{\left(2\pi\right)}^3} \int \frac{\rmd^3 w'}{{\left(2\pi\right)}^3}\, U(\bx,\bx')\, \exp{\left[-i\left(\boldell\cdot \bw + \boldell' \cdot \bw'\right)\right]}.
\end{align}
Combining equation~(\ref{Phi1l_f1l}) with equation~(\ref{f1l}), we can eliminate $\Tilde{f}_{1\boldell}$ to obtain

\begin{widetext}
\begin{align}
\Tilde{\Phi}_{1\boldell}(\bI) = - {\left(2\pi\right)}^3 \sum_{\boldell'} \int \rmd \bI'\, \boldell' \cdot \frac{\partial f_0}{\partial \bI'} \frac{\Psi_{\boldell\boldell'}(\bI,\bI')}{\omega - \boldell'\cdot \Omega'} \left[\Tilde{\Phi}_{1\boldell'}(\bI') + \Tilde{\Phi}_{\rmP\boldell'}(\bI')\right] + {\left(2\pi\right)}^3 i \sum_{\boldell'} \int \rmd \bI' \frac{\Psi_{\boldell\boldell'}(\bI,\bI')}{\omega - \boldell'\cdot \Omega'} f_{1\boldell'}(\bI',0).
\label{Phi1l_implicit}
\end{align}
\end{widetext}
This is an implicit equation for $\Tilde{\Phi}_{1\boldell}$ and thus requires further simplification before a solution is attempted.

\subsection{Bi-orthogonal basis method}

A standard way to solve Equation~(\ref{Phi1l_implicit}) is by expanding the potential and density in the bi-orthogonal basis $(\psi^{(p)},\rho^{(p)})$ that solve the Poisson equation \citep[][]{Kalnajs.77}:

\begin{align}
\Phi_1(\bx,t) &= \sum_{p} a_p(t) \psi^{(p)} (\bx), \;\;\; \Phi_\rmP(\bx,t) = \sum_{p} b_p(t) \psi^{(p)} (\bx) \nonumber\\
\rho_1(\bx,t) &= \sum_{p} a_p(t) \rho^{(p)} (\bx),
\end{align}
such that

\begin{align}
&\psi^{(p)} (\bx) = \int \rmd^3 x'\, U(\bx,\bx') \rho^{(p)} (\bx'), \nonumber\\
&\int \rmd^3 x\, \psi^{(p)\ast}(\bx)\, \rho^{(q)}(\bx) = -4\pi G\, \delta_{pq}.
\end{align}
In this basis, $\Psi_{\boldell\boldell'}(\bI,\bI')$ reduces to

\begin{align}
\Psi_{\boldell\boldell'}(\bI,\bI') = -\frac{1}{4\pi G}\sum_{p} \psi^{(p)}_{\boldell} (\bI) \psi^{(p)\ast}_{\boldell'} (\bI),
\end{align}
where

\begin{align}
\psi^{(p)}_{\boldell} (\bI) = \frac{1}{{\left(2\pi\right)}^3} \int d^3 w\, \psi^{(p)}(\bx) \exp{\left[-i\boldell\cdot \bw\right]}.
\end{align}

In the bi-orthogonal basis, the implicit equation for $\Tilde{\Phi}_{1\boldell}$ given by equation~(\ref{Phi1l_implicit}) reduces to the following matrix equation:

\begin{align}
\Tilde{\mathbb{\ba}}(\omega) = {\left(\varmathbb{I}-\varmathbb{M}(\omega)\right)}^{-1} \left(\mathbb{\bs}(\omega) + \varmathbb{M}(\omega)\, \Tilde{\mathbb{\bb}}(\omega) \right)\,,
\label{lin_resp_sg}
\end{align}
where $\Tilde{\mathbb{\ba}} = \{a_1,a_2,...\}$ is the response vector and $\Tilde{\mathbb{\bb}} = \{b_1,b_2,...\}$ is the perturbation vector. The response matrix $\varmathbb{M}$ is given by

\begin{align}
\varmathbb{M}_{pq} (\omega) = \frac{{\left(2\pi\right)}^3}{4\pi G} \sum_{\boldell} \int \rmd \bI\; \boldell \cdot \frac{\partial f_0}{\partial \bI}\, \frac{\psi^{(p)\ast}_{\boldell}(\bI) \psi^{(q)}_{\boldell}(\bI)}{\omega - \boldell\cdot \Omega}\,.
\label{resp_matrix_app}
\end{align}
The vector corresponding to the initial DF perturbation is given by

\begin{align}
\mathbb{\bs}_p(\omega) = {\left(2\pi\right)}^3 i \sum_{\boldell} \int \rmd \bI\, \frac{f_{1\boldell}(\bI,0)}{\omega - \boldell\cdot \Omega}\, \psi^{(p)\ast}_{\boldell}(\bI).
\end{align}
Note that this assumes the unit of $\psi^{(p)}_{\boldell}$ to be $G/\sqrt{\left|\bx\right|}$ and that of $a_p$ or $b_p$ to be $M/\sqrt{\left|\bx\right|}$ ($M$ is mass). 

\subsection{Temporal response}

The temporal response can be obtained by taking the inverse Laplace transform of equation~(\ref{lin_resp_sg}):

\begin{align}
\mathbb{\ba} (t) &= \frac{1}{2\pi} \int_{ic-\infty}^{ic+\infty} \rmd \omega\, \exp{\left[-i\omega t\right]}\, \Tilde{\mathbb{\ba}} (\omega) \nonumber\\
&= \frac{1}{2\pi} \int_{ic-\infty}^{ic+\infty} \rmd \omega\, \exp{\left[-i\omega t\right]} \nonumber\\
&\times {\left[\varmathbb{I}-\varmathbb{M}(\omega)\right]}^{-1} \left[\mathbb{\bs}(\omega) + \varmathbb{M}(\omega)\, \Tilde{\mathbb{\bb}}(\omega) \right],
\end{align}
where $c$ is chosen such that the integration contour lies in the region of convergence of $\Tilde{\mathbb{\ba}}$. Typically, this means that $c$ exceeds the maximum of the real parts of the poles of $\Tilde{a}_p$. The contribution to the inverse Laplace transform comes from the poles of $\Tilde{\mathbb{\ba}}$, i.e., the poles of $\Tilde{\mathbb{\bb}}$, $\omega = \boldell\cdot \Omega$, and the values of $\omega$ such that

\begin{align}
{\rm det} \left[\varmathbb{I} - \varmathbb{M} (\omega)\right] = 0.
\label{disp_rel}
\end{align}
The discrete values of $\omega$, $\omega_n$, which follow this dispersion relation correspond to the self-sustaining oscillations of the system, known as point modes. All the point modes of a stable self-gravitating system are damped, i.e., have ${\rm Re}(\omega_n)<0$. This phenomenon is known as Landau damping. In an unstable system, one or more of the point modes grows (${\rm Re}(\omega_n)>0$). When a system is marginally stable, the real part of one of the modes sits very close to zero, while all other modes are heavily damped.

The coefficient of the total potential is equal to $\Tilde{\ba}+\Tilde{\bb} = {\left(\varmathbb{I}-\varmathbb{M}\right)}^{-1} \Tilde{\bb}$ (assuming that $f_{1\boldell}\left(\bI,0\right)=0$, i.e., $\bs = 0$). For simplicity, $\bb(t)$ can be expanded as the following Fourier series:

\begin{align}
\bb(t) = \int \rmd\omega^{(\rmP)} \exp{\left[-i\omega^{(\rmP)} t\right]}\, \bb\left(\omega^{(\rmP)}\right),
\end{align}
which can be Laplace transformed to yield

\begin{align}
\Tilde{\bb}\left(\omega\right) = i\int \rmd\omega^{(\rmP)}\,\frac{\bb\left(\omega^{(\rmP)}\right)}{\omega - \omega^{(\rmP)}}.
\end{align}
Now, upon performing the inverse Laplace transform of $\ba + \bb$, we obtain the following temporal dependence for the Fourier mode of the total potential (including the perturber potential and the linear response):

\begin{align}
&\Phi_{\boldell}\left(\bI,t\right) = \Phi_{\rmP\boldell}\left(\bI,t\right) + \Phi_{1\boldell}\left(\bI,t\right) = \left(a_p(t) + b_p(t)\right)\psi^{(p)}_{\boldell}\left(\bI\right) \nonumber\\
&= \int \rmd\omega^{(\rmP)} \exp{\left[-i\omega^{(\rmP)} t\right]}\, {\left[\varmathbb{I}-\varmathbb{M}\left(\omega^{(\rmP)}\right)\right]}^{-1}_{pq} \, b_q\left(\omega^{(\rmP)}\right) \psi_{\boldell}^{(p)}\left(\bI\right),
\label{Phil_t}
\end{align}
where we have taken the long time limit, i.e., evaluated the response at times longer than the damping time of the least damped Landau mode, assuming that the system is linearly stable. 

The linear response in the DF can be obtained by taking the inverse Laplace transform of $f_{1\boldell}$ from equation~(\ref{f1l}):

\begin{widetext}
\begin{align}
f_{1\boldell}\left(\bI,t\right) = -\boldell\cdot\frac{\partial f_0}{\partial \bI} \int \rmd\omega^{(\rmP)} \frac{b_q\left(\omega^{(\rmP)}\right) \psi_{\boldell}^{(p)}\left(\bI\right)}{\omega^{(\rmP)}-\boldell\cdot{\bf\Omega}}\left[{\left(\varmathbb{I}-\varmathbb{M}\left(\omega^{(\rmP)}\right)\right)}^{-1}_{pq}\exp{\left[-i\omega^{(\rmP)}t\right]}-{\left(\varmathbb{I}-\varmathbb{M}\left(\boldell\cdot{\bf\Omega}\right)\right)}^{-1}_{pq}\exp{\left[-i\boldell\cdot{\bf\Omega}t\right]}\right].
\label{f1l_t}
\end{align}
\end{widetext}
The response thus consists of a term that follows the temporal dependence of the perturber and another that oscillates at the unperturbed frequencies but is dressed by collective interactions.

\section{Quasilinear response theory}\label{App:QL_resp}

Linear response theory describes the evolution of the fluctuations on top of a smooth background, but the background itself evolves due to the combined action of the linear fluctuations. Modeling this requires performing a second order or quasilinear perturbation of the Vlasov-Poisson equations. The second order response equation for the Fourier transform of $f_2$ is given by

\begin{align}
\frac{\partial f_{2\boldell}}{\partial t} + i\boldell\cdot{\bf \Omega}f_{2\boldell} &= i\boldell\cdot\frac{\partial f_0}{\partial \bI}\Phi_{2\boldell} \nonumber\\
& + i\sum_{\boldell'} \left[\boldell'\cdot\frac{\partial f_{1\boldell-\boldell'}}{\partial\bI} \left(\Phi_{1\boldell'}+\Phi_{\rmP\boldell'}\right)\right.\nonumber\\
&\left.-\left(\boldell-\boldell'\right)\cdot \frac{\partial \left(\Phi_{1\boldell'}+\Phi_{\rmP\boldell'}\right)}{\partial \bI}f_{1\boldell-\boldell'}\right].
\label{2nd_CBE_lmode}
\end{align}
The evolution of the phase-averaged DF, $\int \rmd^3w\, f_2 /{\left(2\pi\right)}^3 = f_{2\boldell\to 0} = f_0$, is obtained by putting $\boldell=0$ in the above equation, and is given by the following quasilinear equation:

\begin{align}
\frac{\partial f_{0}}{\partial t} = i\sum_{\boldell}\boldell\cdot\frac{\partial}{\partial \bI}\left<f^\ast_{1\boldell}\left(\bI,t\right)\Phi_{\boldell}\left(\bI,t\right)\right>,
\label{QL_eq_app}
\end{align}
where we have defined $\Phi_{\boldell} = \Phi_{\rmP\boldell} + \Phi_{1\boldell}$, used the reality condition that $f_{1,-\boldell} = f_{1\boldell}^\ast$, and absorbed the factor $\epsilon^2$ in the correlation in the RHS. The brackets $\left<Q\right>$ denote the ensemble average of the quantity $Q$ over random phases. 

Now we assume that the perturber potential assumes the following form of a red noise:

\begin{align}
\left<b^\ast_q\left(t\right) b_{q'}\left(t'\right)\right> = B^\ast_q B_{q'} \calC_t\left(t-t'\right),
\end{align}
where $\calC_t$ denotes the temporal correlation function, which is equal to $\delta\left(t-t'\right)$ for white/uncorrelated noise. Therefore, the Fourier transform of $b_q(t)$, $b_q\left(\omega^{(\rmP)}\right)$, follows the condition:

\begin{widetext}
\begin{align}
\left<b_q^\ast\left(\omega^{(\rmP)}\right)b_{q'}\left(\omega^{(\rmP)}\right)\right> &= \frac{1}{{\left(2\pi\right)}^2} \int \rmd t \int \rmd t' \exp{\left[i\left(\omega^{(\rmP)}t - \omega^{'(\rmP)}t'\right)\right]} \left<b^\ast_q\left(t\right) b_{q'}\left(t'\right)\right> \nonumber\\
&= B^\ast_q B_{q'}\, \calC_{\omega}\left(\omega^{(\rmP)}\right) \delta\left(\omega^{(\rmP)}-\omega^{'(\rmP)}\right),
\end{align}
\end{widetext}
where $\calC_\omega$ denotes the Fourier transform of $\calC_t$.

Substituting the linear responses from equations~(\ref{f1l_t}) and (\ref{Phil_t}) in the quasilinear equation~(\ref{QL_eq_app}), we obtain

\begin{align}
\frac{\partial f_0}{\partial t} = \sum_{\boldell}\boldell\cdot\frac{\partial}{\partial \bI} \left(D_{\boldell}\left(\bI,t\right) \, \boldell\cdot\frac{\partial f_0}{\partial \bI}\right),
\end{align}
where $D_{\boldell}\left(\bI,t\right)$ is given by

\begin{widetext}
\begin{align}
D_{\boldell}\left(\bI,t\right) &= -i\int \rmd \omega^{(\rmP)} \, \calC_{\omega}\left(\omega^{(\rmP)}\right) \frac{B^\ast_q\left(\omega^{(\rmP)}\right)B_{q'}\left(\omega^{(\rmP)}\right) \psi^{(p)\ast}_{\boldell} \psi^{(p')}_{\boldell}}{\omega^{(\rmP)}-\boldell\cdot{\bf\Omega}} {\left(\varmathbb{I}-\varmathbb{M}\left(\omega^{(\rmP)}\right)\right)}^{-1}_{pq} \nonumber\\
&\times\left[{\left(\varmathbb{I}-\varmathbb{M}^\ast\left(\omega^{(\rmP)}\right)\right)}^{-1}_{p'q'} - {\left(\varmathbb{I}-\varmathbb{M}^\ast\left(\boldell\cdot{\bf\Omega}\right)\right)}^{-1}_{p'q'}\exp{\left[-i\left(\omega^{(\rmP)}-\boldell\cdot{\bf\Omega}\right)t\right]}\right].
\end{align}
\end{widetext}
In the long time limit, which is what we are interested in, $D_{\boldell}\left(\bI,t\right)$ reduces to

\begin{align}
&\lim_{t\to\infty} D_{\boldell}\left(\bI,t\right) = D_{\boldell}\left(\bI\right)\nonumber\\
&= {\left|{\left(\varmathbb{I}-\varmathbb{M}\left(\boldell\cdot\bf{\Omega}\right)\right)}^{-1}_{pq} B_q \psi_{\boldell}^{(p)}\left(\bI\right)\right|}^2 \calC_{\omega}\left(\boldell\cdot{\bf\Omega}\right).
\end{align}
Here we have used the identity that $\lim_{t\to\infty}\exp{\left[-ixt\right]}/x = 1/x - i\pi\delta\left(x\right)$ with $x = \omega^{(\rmP)}-\boldell\cdot{\bf\Omega}$.

%\nocite{*}

\bibliography{references_banik}% Produces the bibliography via BibTeX.

\end{document}